\documentclass[12pt]{article}

\usepackage{mathptmx}
\usepackage{amssymb,amsmath,amsbsy,amsfonts}
\usepackage{graphicx}
\usepackage[left=2cm,right=2cm,top=2cm,bottom=2cm]{geometry}
\usepackage{braket}
\usepackage{color}
\usepackage{mathrsfs}
\usepackage{cite}
\usepackage{physics}
\usepackage{subfig}
\usepackage{float}
\usepackage{hyperref}
\hypersetup{colorlinks = true,
	linkcolor = red,
	anchorcolor = blue,
	citecolor = blue,
	filecolor = blue,
	urlcolor = blue}

\numberwithin{equation}{section}
	
\begin{document}

\title{Exotic light transition between superluminal and subluminal group velocity in a $\mathcal{PT}$ coupled slab waveguide.}

\author{B.M. Villegas-Martínez$^*$, H.M. Moya-Cessa, F. Soto-Eguibar\\
\small Instituto Nacional de Astrofísica, Óptica y Electrónica, INAOE \\
\small {Calle Luis Enrique Erro 1, Santa María Tonantzintla, Puebla, 72840 Mexico}\\
$^*$\small {Corresponding author: bvillegas@inaoep.mx}\\}
	
\date{\today}

\maketitle

\begin{abstract} 
We propose a rigorous full-wave modal analysis based on the eigenmodes approach (\textquotedblleft supermodes\textquotedblright), that allows the calculation of the transverse electric ($TE$) and transverse magnetic ($TM$) stationary propagation eigenmodes in a coupled $\mathcal{PT}$-symmetric slab waveguide. Our findings show that the group velocity, for the fundamental $TE_{0}$ supermode, passes from superluminal to subluminal speed over both sides of a critical point (CP), located near to the exceptional point (EP). We also demonstrate that the first higher-order supermode, $TE_{1}$, exhibits a slow light phenomenon around the EP.
\end{abstract}

\section{Introduction}
In recent years, the subject of controlling the subluminal (slow light) and superluminal (fast light) group velocities in structures and dispersive media have captured strong growing attention in the photonic scientific and engineering communities. Various mechanisms have been developed to investigate and control both phenomena. Slow light has been observed from its origins in atomic vapours by self-induced transparency\cite{1,2,3,4}. Shortly thereafter, group velocity reductions were achieved in quantum dots at room temperature\cite{5,6,7}, Bose–Einstein condensates\cite{8}, and semiconductors operating at telecommunication wavelengths\cite{9}. Conversely, fast light has been observed in gaseous systems\cite{10,11,12,13}, theoretical reported in a rubidium atomic system\cite{14}, in coupled optical resonators\cite{15}, and followed by a demonstration in an alexandrite crystal\cite{16}. Both slow light and fast light regimes have also been observed in optical fibers\cite{17}, cold atoms\cite{18}, and other mediums \cite{19,20}. These findings strengthened the study and formed an excellent portfolio that hold promise for potential applications of fast and slow light effects in nanoscale technologies and its possible groundbreaking on information processing and communication networks. In this regard, it has prompted an extensive search for new alternative materials of low-cost and high-performance periodic photonic structures, especially for achieving slow or fast group propagation of light.\\
At the same time, the advent of optical devices technology brought a new class of synthetic materials, based on parity‐time ($\mathcal{PT}$) symmetry  \cite{21,22,23,24,25,26,27,28,29}. 
This technology in its standard configuration, known as $\mathcal{PT}$-directional coupler, is composed of two waveguide elements with absorption and amplification into the waveguide refractive index profiles\cite{30,31,32,33,34}, although other suitable $\mathcal{PT}$-symmetric setups can be modeled and developed
\cite{35,36,37,38,39,40}. Aside, the overwhelming interest in these artificial systems, specially in slow light science, has been prompted by the existence of a certain turning point that separates complex modes from purely real ones; such threshold point in $\mathcal{PT}$-photonic systems, also known as exceptional point (EP) \cite{41,42,43,44,45,46,47}, is a direct outcome from the non-Hermitian topological properties of the bicentric structures balanced with gain and loss. A most recent, and cutting-edge example exploiting this direction, was made by Moiseyev et al \cite{48}; they analyzed theoretically and numerically that in a waveguide structure with an asymmetric gain/loss profile, the coalescence of two $TE$ modes, near an EP, makes its group velocity drop to zero. More recently, and in contrast with the previous finding, Lingxuan and coworkers \cite{49} identified a critical point where the dispersion relation of one of the $TE$ modes experiences fast light propagation in the forward and backward directions, and which is different from the EP previously reported by Moiseyev. However, it should be noted that in the stopped light work, the theoretical derivation for the group velocity of the $TE$ modes was obtained by applying a perturbative analysis on their wave equation, and the full-stop of the light pulse was possible by varying the gain-loss parameter in time; the coupled $TE$ guided modes were calculated by diagonalizing the matrix representation in a sine basis with a finite window in their simulation protocol, where the background index was discarded and the EP was outside of the operation window; in particular, the Hamiltonian becomes non-diagonalizable at the EP. The same situation occurs for the results of the coupled $\mathcal{PT}$-symmetric waveguides reported by Lingxuan, where the solutions were obtained only in the $\mathcal{PT}$-symmetric phase. However, in both cases, the matrix representation of the non-Hermitian Hamiltonian to get the propagation constants was found by using the coupled-mode theory; although coupled-mode theory offers a simple and valuable physical insight into how a $\mathcal{PT}$-symmetric pair of waveguides are coupled \cite{50,51,52}, such a procedure has several limitations. For instance, the formalism assumes that the waveguides are identical, and weakly $\mathcal{PT}$-symmetric (i.e., that the perturbation that breaks the $\mathcal{PT}$-symmetry is small); as a result, it may not be able to provide a complete picture of the dynamics of the system. These limitations make it difficult to apply the theory to a wider range of systems, including those that are strong $\mathcal{PT}$-symmetric, where the transverse profiles of the modes are affected by the coupling, or structures made up of various types of waveguides, where the propagation constants of the modes may be different and should be computed by using a different analysis. Therefore, in this work, we seek to supplement Moiseyev's research, as well as Lingxuan's study, with an alternative route that meets both the weakly and strongly guiding conditions to accurately describe the $\mathcal{PT}$-optical system. The method presented in this contribution performs a full-wave modal analysis of the $\mathcal{PT}$ waveguide structure, based on the eigenmodes approach, a technique similar to the procedure used by Nolting et al \cite{53} to study a strongly $\mathcal{PT}$-symmetric system. The concept of the eigenmode approach relies on the doctrine of the interference pattern of supermodes, which is another popular approach used to study the dynamics of coupled systems; specifically, the formalism allows a more global understanding and deeper insight into the characteristics of the system, by looking at the behavior of the supermodes as a whole. Moreover, it can be used to show how the system behaves as it transitions from unbroken to broken $\mathcal{PT}$- regimens, without the use of operation windows to discard the EP. Moreover, it also permits to analyze the structure when the gain and losses are strong. Thus, in this work, the supermode theory is applied to the general analysis of a couple $\mathcal{PT}$-symmetric slab waveguide. Here, the non-Hermitian system supports two coexisting supermodes, each of them will count with a complex propagation constant, which can be decomposed into two eigenstates of polarization: transverse electric($TE$) or transverse magnetic ($TM$), with an explicit gain or loss characterization. The analysis of their optical wave propagation leads to a transcendental eigenequation, where one or several intrinsic parameters of the $\mathcal{PT}$-waveguide system can be varied. The eigenequation allows us to determine the $TE/TM$ propagation constants for the entire structure and can be used to study the different degrees of imperfection in a practical device. We revealed from our analysis that all the essential features of the system can be extracted from its supermodes propagation constants, where the simulation results show that the solutions are in agreement with those obtained from references\cite{48,49} in the vicinity of the EP. As a consequence, this allows us to determine numerically the dispersion diagram of the group velocity in the unbroken and broken $\mathcal{PT}$-regimen, for two $TE$ supermodes, as a function of the angular frequency. We show that the light over the waveguide system, in the fundamental supermode, undergoes an abrupt superluminal-slow transition near the EP. Furthermore, we show that the first higher-order supermode exhibits a subluminal group velocity. The non-exactly-$\mathcal{PT}$-symmetric configuration, consisting of a geometric asymmetry of unequal slab widths, is also investigated. Leading to a situation where both light phenomena are not conserved when the width mismatching between the two slabs is increased. \\
Accordingly, the rest of the work is laid out as follows: In Section 2, we present the full-wave modal analysis in the physical waveguide configuration under study. Next, Section 3, contains our main results; we show the dispersion diagram for the fundamental $TE_{0}$ supermode and the first higher-order $TE_{1}$ supermode, when the balance of gain and loss is assured; we perform a numerical analysis of the group velocity of each supermode. Section 4 goes into details for the unbalanced model, when the absorbing and amplifying regions do not have the same thickness, demonstrating the detuning near an EP. Finally, the conclusions are stated in Section 5.

\section{Two slab waveguide model and solution}
\begin{figure}[H]
    \centering
	{\includegraphics[width=0.45\textwidth]{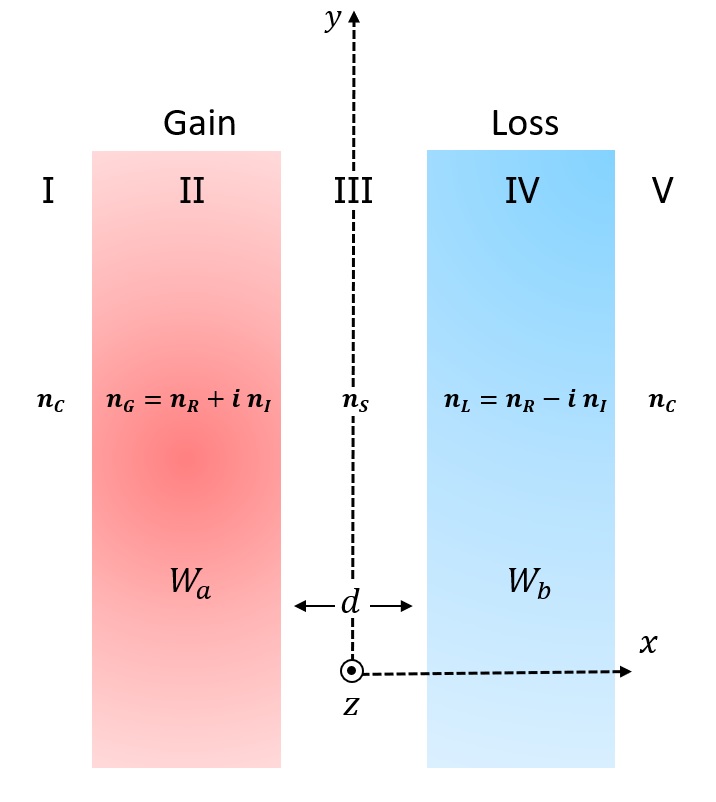}}
	\caption{Schematic representation of two coupled slab waveguides with gain (red, left) and loss (blue, right). The quantities $W_{a}$ and $W_{b}$ represent the slab thickness, $n_{C}$ is the cladding index, $n_{S}$ is a centered substrate, and the sign of the imaginary part of the refractive indexes, $n_{G}=n_{R}+i n_{I}$ and $n_{L}=n_{R}-i n_{I}$, represent gain and losses. The light propagates along the $z$-direction and there is no field variation in the $y$-direction.}
	\label{f1}
\end{figure}

We set the stage with the description of the physical waveguide configuration under study. The reference model system consists of a pair of parallel coupled slabs waveguides with thickness $W_{a}$ and $W_{b}$ separated by a gap $d$. Such a structure can be considered as a five-layer optical waveguide as seen in Fig.\ref{f1}; the slab which occupies region II has a refractive index $n_{G}=n_{R}+i n_{I}$, whereas the refractive index of slab on region IV is given by $n_{L}=n_{R}- i n_{I}$. Here, $n_{R}$ stands for the real index part, and $n_{I}=\frac{\lambda}{2 \pi} \alpha$ denotes the imaginary part, which depends on the vacuum wavelength $\lambda$ and the gain-loss parameter $\alpha$; similar to Ref.\cite{48}, the index of refraction is frequency dependent because the wavelength is equal to $2 \pi/k$ with $k = \omega/c$, where $\omega$ denotes the angular frequency of the optical field and $c$ the speed of light. Moreover, the negative and positive imaginary parts of the refractive index denote the gain or lossy medium in the guiding layers. Between the two slabs, there is a substrate with refractive index, $n_{S}$, and two clad layers, $n_{C}$, are located on opposite sides in the waveguide structure. In our model, the refractive indexes are chosen in a way that the condition $n_{R}>n_{C}\geq n_{S}$ is satisfied, in order that the propagation takes place in the slab’s region. A possible realization of the system studied can be made by considering ZnSiAs2, GaAs or InGaAsP as the guiding material, where the refractive index variation can be induced by femtosecond-scale all-optical switching\cite{54}. It is important to point out, that under the geometric constraint that both guides have equal widths ($W_{a} = W_{b}$), an ideal $\mathcal{PT}$-symmetric configuration is achieved; in this configuration, the complex refractive index can be assimilated as a complex potential that satisfies the essential condition $n\left(x\right)=n^{*}\left(-x\right)$, when the reference system is placed at the center of the $x$ axis.
Generally, if we consider a light beam propagating along the $z$ axis and the transverse direction of the field distribution is along the $x$-axis, with no variation of fields in the $y$-direction, the expression for the electromagnetic field, 
$\left(
\begin{matrix}
E\left(x,y,z,t\right)  \\
H\left(x,y,z,t\right) 
\end{matrix}
\right)=\left(
\begin{matrix}
E_{y}\left(x\right)  \\
H_{y}\left(x\right) 
\end{matrix}
\right) e^{i\left(\omega t-\beta z\right)}$, can be decomposed in terms of the transverse electric ($TE$) and transverse magnetic ($TM$) modes, which satisfy the one-dimensional Helmholtz equation at each interface of the five regions
\begin{equation} \label{1}
\left[\frac{\partial^2}{\partial x^2} + k^2 n^2_{j}\right]\phi_{j,y}(x)=\beta^2\phi_{j,y}(x),
\end{equation}
where $\phi_{j,y}(x)$ denotes $E_{j,y}(x)$ for the $TE$ modes or $H_{j,y}(x)$ for the $TM$ modes, and the subindex $j=G,L,C,S$ represents gain, loss, clad and substrate regions of space, respectively. Like in a conventional passive coupled slab waveguide, $\beta$ represents the modal propagation constant and $n_{j}$ is the refractive index in the $j$-th layer. Evidently, Eq.\eqref{1} is analogous to the one-dimensional stationary Schrödinger equation:$\left[-\frac{1}{2}\frac{\partial^2}{\partial x^2} + V\left(x\right)\right]\Psi\left(x\right)=\Psi\left(x\right)E$, with potential $V\left(x\right)=-\frac{1}{2}k^2 n^2_{j}$ and energy eigenvalue $E=-\frac{1}{2}\beta^2$, where the propagating modes are equal to bound states \cite{55}. Then, the quantum scenario analogous to the $\mathcal{PT}$-symmetric waveguide system is a configuration of a finite double-well with a complex potential. \\
Since we are interested in light that propagates inside of the gain and loss channels, exponentially decays in the cladding and the interference between both modal amplitudes take place on the substrate. Then, the stationary modes with transverse profiles, $\phi_{y}\left(x\right)$, frequency $\omega$ and propagation constant $\beta$, can be taken to be
\begin{equation} \label{2}
\phi_{y}\left(x\right) = \left\{
	       \begin{array}{lll}
		 I:   & A_{1} e^{\gamma_{C}\left(x+a\right)}  & x< -a \\
		 II:  & A_{2} \cos\left[k_{G}\left(x+d/2\right)\right]+A_{3} \sin\left[k_{G}\left(x+d/2\right)\right] & -a< x<-d/2\\
		 III:    & A_{4} e^{-\gamma_{S}x} + A_{5} e^{\gamma_{S}x} & -d/2<x<d/2 \\
		 IV: &  A_{6} \cos\left[k_{L}\left(x-d/2\right)\right]+A_{7} \sin\left[k_{L}\left(x-d/2\right)\right] & d/2<x<b\\
		 V: & A_{8} e^{-\gamma_{C}\left(x-b\right)} & x>b
	       \end{array}
	     \right. ,
   \end{equation}
being $a=W_{a}+d/2$, $b=W_{b}+d/2$ and $A_{n}$, with $n=1,2..8$, denotes the unknown amplitude coefficients of the field. The quantities $k^2_{G}=k^2 n^2_{G}-\beta^2$, $k^2_{L}=k^2 n^2_{L}-\beta^2$, $\gamma^2_{S}=\beta^2-k^2 n^2_{S}$ and $\gamma^2_{C}=\beta^2-k^2 n^2_{C}$ correspond to the transverse field parameters in the gain, loss, substrate and cladding regions, respectively. It is our purpose to find the modal propagation constant of the optical field in the above full waveguide system, given the geometrical dimensions of each layer and their refractive indexes. Hence, solving equation \eqref{1} in each layer and imposing the continuity and boundary conditions for the transverse electric field $\phi_{j,y}\left(x\right)=\phi_{m,y}\left(x\right)$, $\frac{d}{dx}\phi_{j,y}\left(x\right)=\frac{d}{dx}\phi_{m,y}\left(x\right)$, and the transverse magnetic field $\frac{d}{dx}\phi_{j,y}\left(x\right)=\left(\frac{n_{j}}{n_{m}}\right)^2\frac{d}{dx}\phi_{m,y}\left(x\right)$, at the interfaces between the $j$-th and $m$-th layer, lead to a system of eight equations  for the amplitudes $A_{n}$ (which is relegated to Appendix A) that combined yield a single equation, in terms of the propagation constant $\beta$, which is
\begin{equation} \label{3}
\varphi \left(\beta,\omega\right)=e^{\gamma_{S}d}H^{+}_{G} H^{+}_{L}-e^{-\gamma_{S}d}H^{-}_{G} H^{-}_{L}=0,
\end{equation}
with
\begin{align} \label{4}
H^{\pm}_{L}=& \cos\left( k_{L} W_{b} \right) \left[\left(\frac{\gamma_{S}}{\gamma_{C}}\right) \left(\frac{n_{C}}{n_{S}}\right)^{2p} \pm 1 \right] +  \sin\left( k_{L} W_{b} \right) \left[\left(\frac{\gamma_{S}}{k_{L}}\right) \left(\frac{n_{L}}{n_{S}}\right)^{2p} \mp \left(\frac{k_{L}}{\gamma_{C}}\right) \left(\frac{n_{C}}{n_{L}}\right)^{2p} \right], \nonumber \\
H^{\pm}_{G}=& \cos\left(k_{G} W_{a}\right)\left[ \left(\frac{\gamma_{S}}{\gamma_{C}}\right) \left(\frac{n_{C}}{n_{S}}\right)^{2p} \pm 1 \right] +  \sin\left(k_{G} W_{a}\right) \left[\left(\frac{\gamma_{S}}{k_{G}}\right) \left(\frac{n_{G}}{n_{S}}\right)^{2p} \mp \left(\frac{k_{G}}{\gamma_{C}}\right) \left(\frac{n_{C}}{n_{G}}\right)^{2p} \right].
\end{align}
This transcendental equation, known as dispersion equation, enables one to determine the propagation constant $\beta$ for the $TE$ (when $p=0$) or $TM$ (for $p=1$) modes at a given operation frequency $\omega$. It is important to point out that the transverse field parameters $k_{G}$, $k_{L}$, $\gamma_{S}$ and $\gamma_{C}$ are all functions of $\beta$, therefore Eq.\eqref{3} is a single variable function of $\beta$. Then, if one specifies all slab waveguide parameters, we can numerically determine the roots of the dispersion equation in the complex plane for all possible solutions of $\real\left(\beta_{TE/TM}\right)$. It should be kept in mind, that the roots we are looking for are the eigenvalues of the transcendental equation, under the cut-off frequency of the modes $\frac{\omega n_{S}}{c}< \real\left(\beta_{TE/TM}\right)<\frac{\omega n_{R}}{c}$, where the sign of $\beta_{TE/TM}$ has been chosen as $\real\left(\beta_{TE/TM}\right)>0$. 

\subsection{Slab waveguide with distributed gain and loss: case when $W_{a} = W_{b}$}
\begin{figure}[H]
    \centering
	\subfloat[]
	{\includegraphics[width=0.48\textwidth]{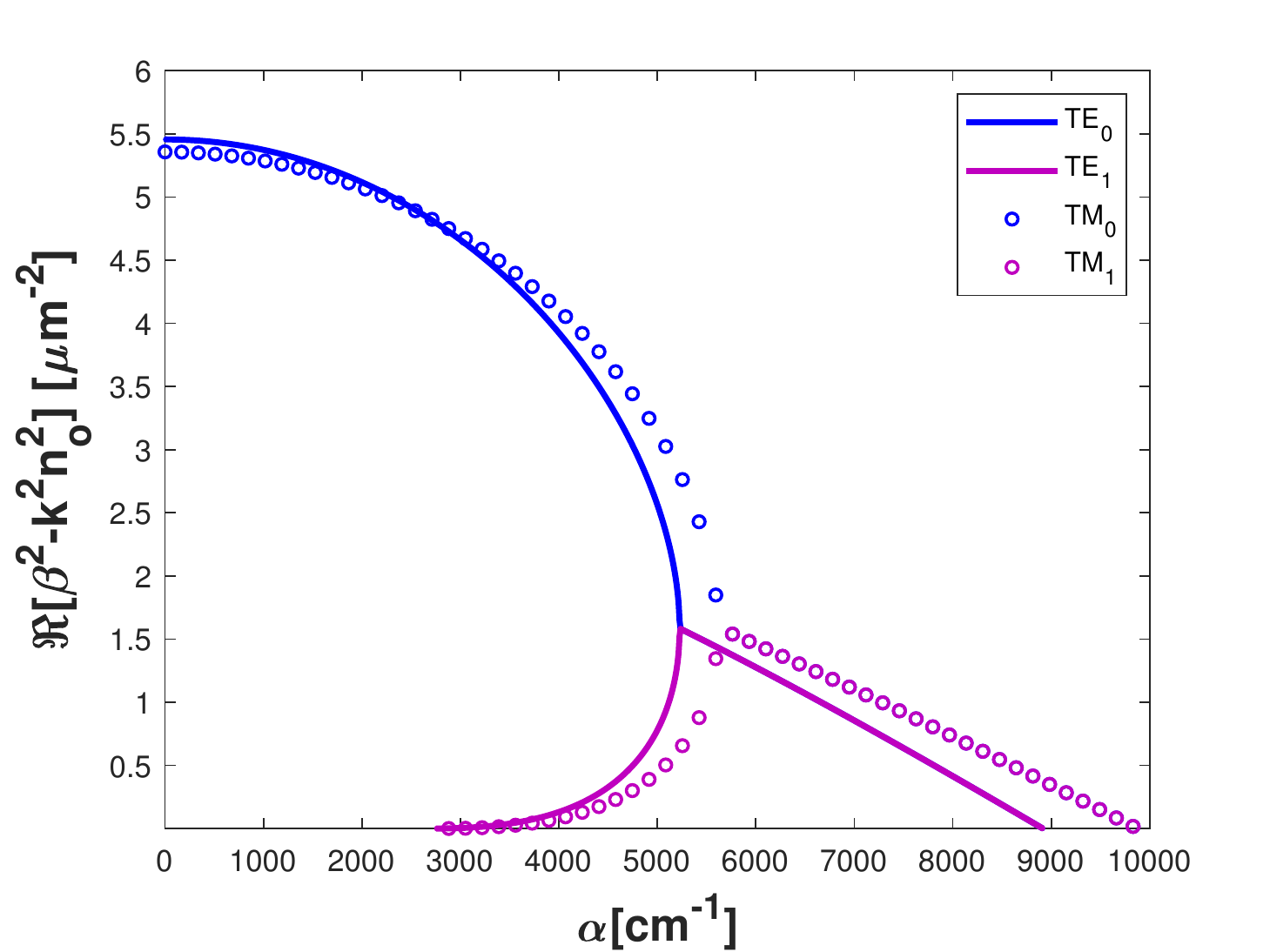}}
	\quad
	\subfloat[]
	{\includegraphics[width=0.48\textwidth]{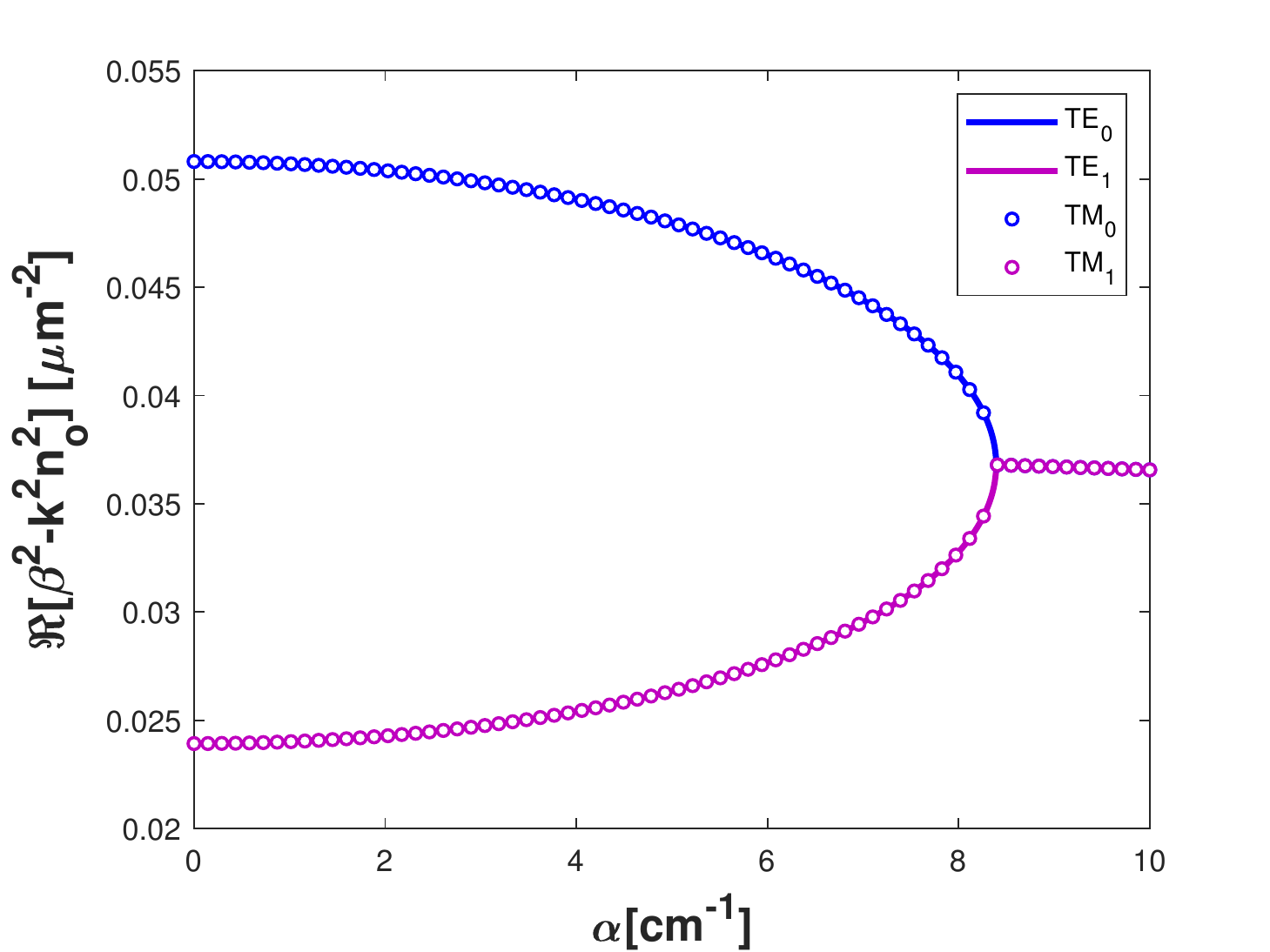}}
	\caption{Propagation constants of $TE$ modes and $TM$ modes as function of the gain-loss parameter $\alpha$. We have considered that the slab structures operate at $\lambda=1.55 \, \mu\mathrm{m}$ and take into account the system properties for a strongly guiding two-mode structure. (a) $d=0$, $W=0.5 \, \mu\mathrm{m}$, $n_{R}=3.252398$, $n_{o}=3.169355$ and for a weakly guiding two-mode (b) $d=W=5 \, \mu\mathrm{m}$, $n_{R}=3.301$, $n_{o}=3.3$. The figure on the left was obtained by considering $n_{I}=\frac{\lambda}{4 \pi} \alpha$ instead of $n_{I}=\frac{\lambda}{2 \pi} \alpha$ into the numerical analysis. A root-finding method was applied in Eq.\eqref{5} to find 1000 roots separately for each $TE$ supermodes and 60 roots for each $TM$ supermodes.}
	\label{f2}
\end{figure}

\begin{figure}[H]
    \centering
	\subfloat[]
	{\includegraphics[width=0.48\textwidth]{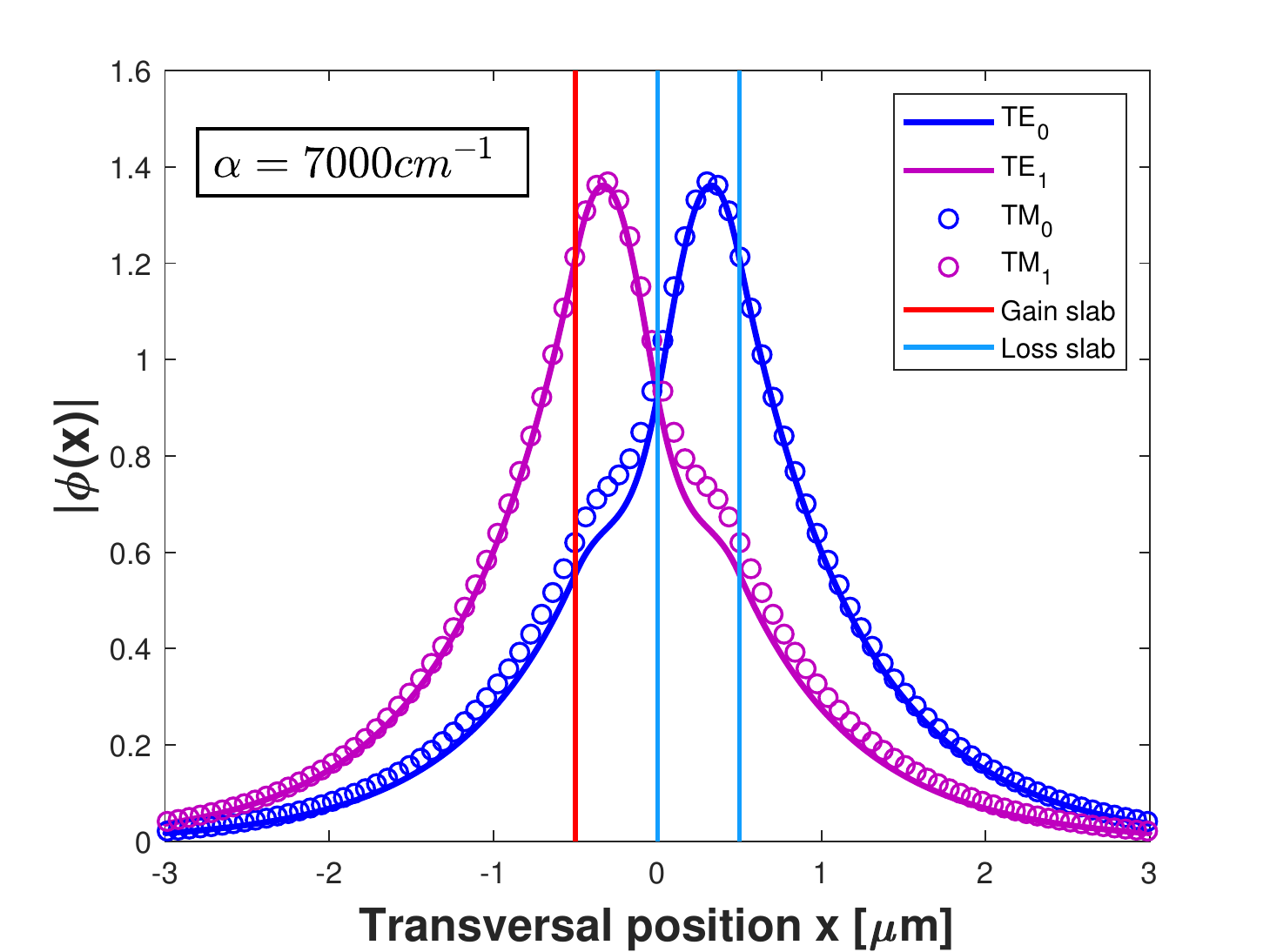}}
	\quad
	\subfloat[]
	{\includegraphics[width=0.48\textwidth]{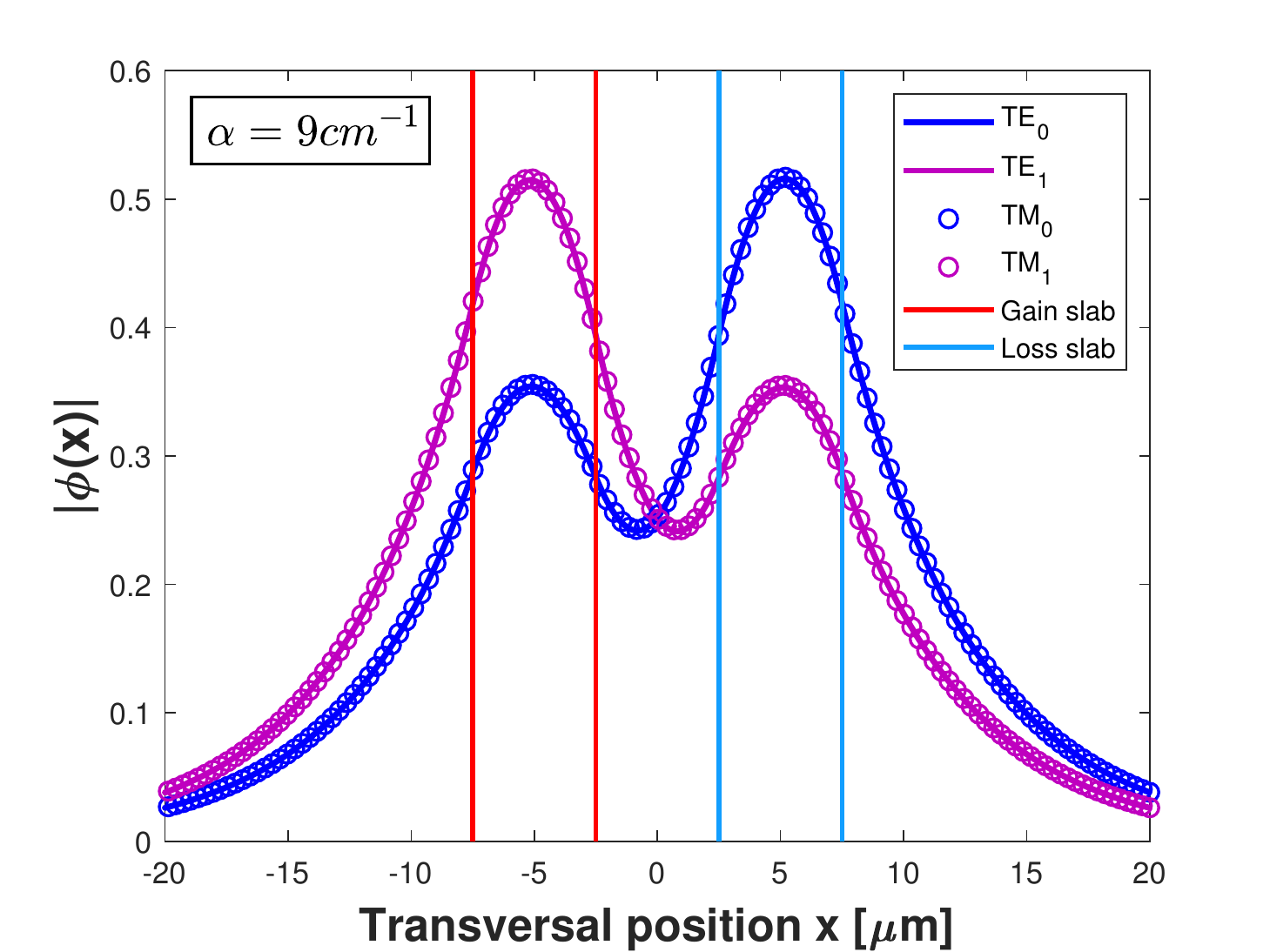}}
	\caption{Electric and magnetic field distributions of $TE$ and $TM$ polarized guided modes in the gain/loss slab waveguide. (a) Strongly guiding two-mode  waveguides with gain-loss parameter $\alpha=7000 \, \mathrm{cm}^{-1}$ and propagation constants: $\beta_{TE_{0,1}}=\left(12.8812 \pm 0.1146i\right)\, \mu\mathrm{m}^{-1} $, $\beta_{TM_{0,1}}=\left(12.8906 \pm 0.0921i\right)\, \mu\mathrm{m}^{-1}$. (b) Weakly guiding two-mode waveguides with gain-loss parameter $\alpha=9 \, \mathrm{cm}^{-1}$ and propagation constants: $\beta_{TE_{0,1}}=\left(13.3785 \pm 0.0002i \right)\, \mu\mathrm{m}^{-1}$, $\beta_{TM_{0,1}}=\left(  13.3785 \pm 0.0002i \right)\, \mu\mathrm{m}^{-1}$. In both waveguide configurations, the field distributions of the $TE/TM$ modes are mutually symmetrically with regard to the $x$ plane.}
		\label{f3}
\end{figure}

\begin{figure}[H]
    \centering
    	\subfloat[]
	{\includegraphics[width=0.30\textwidth]{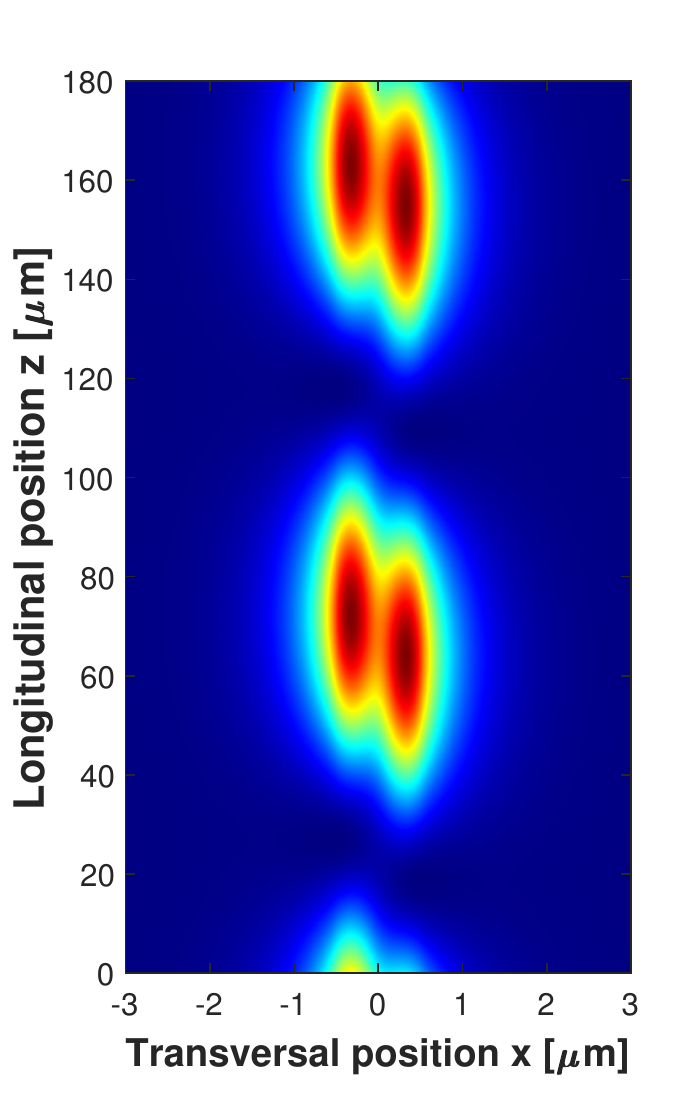}}
		\quad
	\subfloat[]
	{\includegraphics[width=0.30\textwidth]{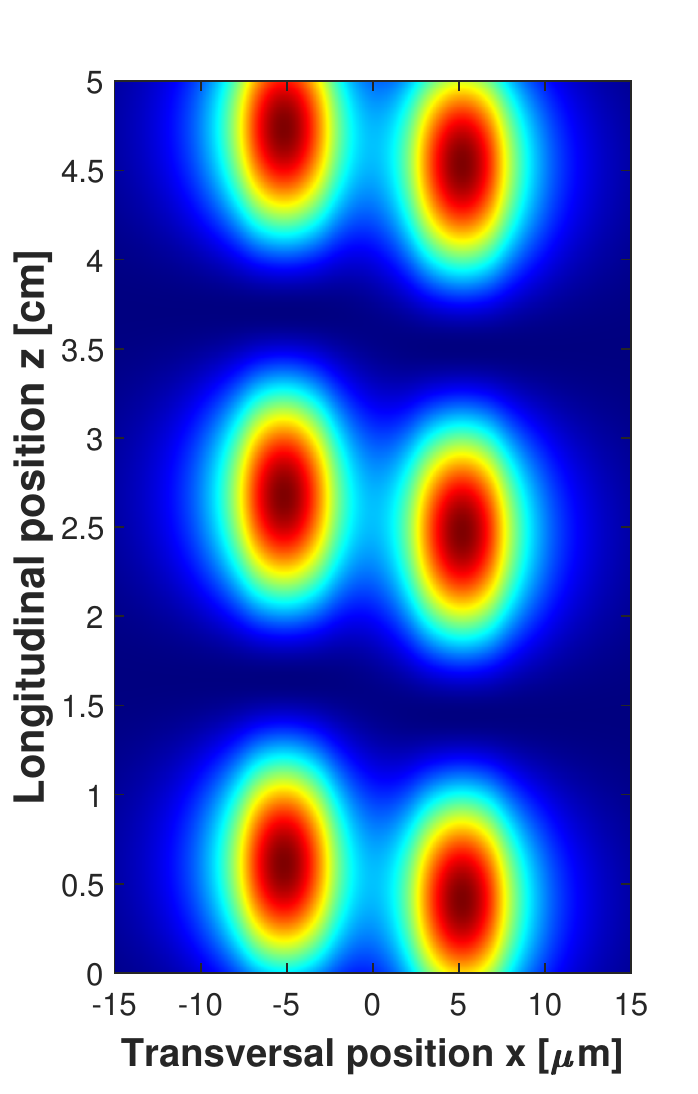}}
	\caption{Power distribution for the propagating total field consisting of the two guided modes. (a) strongly coupled case with gain-loss parameter $\alpha=5000 \, \mathrm{cm}^{-1}$ and with propagation constants: $\beta_{TE_{0}}=12.9469 \, \mu\mathrm{m}^{-1}$ and $\beta_{TE_{1}}=12.8778\, \mu\mathrm{m}^{-1}$. (b) Weakly coupled scenario with gain-loss parameter $\alpha=8 \, \mathrm{cm}^{-1}$ and with propagation constants: $\beta_{TE_{0}}=13.3786 \, \mu\mathrm{m}^{-1}$, $\beta_{TE_{1}}=13.3783 \, \mu\mathrm{m}^{-1}$.}
	\label{f4}
\end{figure}

Equation \eqref{3} describes a general model which is many-layered, in both senses of the word; for instance, it can be used to analyze a range of complex scenarios in a practical $\mathcal{PT}$-slab waveguide system; this includes the study of wave propagation within the structure, when it is placed in various background media, cladding layers, imbalanced gain and loss, and different widths of slab waveguides. Additionally, it can also be used to investigate many transverse electric and magnetic modes (see Appendix B) and their hybrid forms, within the system. Nonetheless, this aspect is not covered in the present article, and rather we focus only on the scenario where the waveguide supports two guided modes for both $TE$ and $TM$ polarizations. In this case, the first stage of our analysis is the simple scenario where two slab waveguides are on a background material with index $n_{o}=n_{C}=n_{S}$, which means the cladding and substrate regions have the same refractive index. Hence, the transverse field parameter in the background medium is given by $\gamma^2_{o}=\beta^2-k^2 n^2_{o}$; moreover, we allow the width of both slabs to be the same, $W_{a}=W_{b}=W$, to maintain a geometrically symmetric structure, with gain and loss balance. Then, the dispersion relationship \eqref{3} is transformed to
\begin{align} \label{5}
\varphi \left(\beta,\omega\right)=&-e^{-\gamma_{o}d} \sin\left(k_{L} W\right) \sin\left(k_{G} W\right) \left[\left(\frac{\gamma_{o}}{k_{G}}\right) \left(\frac{n_{G}}{n_{o}}\right)^{2p} + \left(\frac{k_{G}}{\gamma_{o}}\right) \left(\frac{n_{o}}{n_{G}}\right)^{2p}\right]\left[\left(\frac{k_{L}}{\gamma_{o}}\right) \left(\frac{n_{o}}{n_{L}}\right)^{2p} + \left(\frac{\gamma_{o}}{k_{L}}\right) \left(\frac{n_{L}}{n_{o}}\right)^{2p}\right] \nonumber\\
& + e^{\gamma_{o} d} \Bigg\lbrace 2 \cos\left(k_{L} W\right) + \sin\left(k_{L} W\right) \left[\left(\frac{\gamma_{o}}{k_{L}}\right) \left(\frac{n_{L}}{n_{o}}\right)^{2p} - \left(\frac{k_{L}}{\gamma_{o}}\right) \left(\frac{n_{o}}{n_{L}}\right)^{2p}\right] \Bigg\rbrace \Bigg\lbrace 2 \cos\left(k_{G} W\right) +  \sin\left(k_{G} W\right) \nonumber \\
& \times \left[\left(\frac{\gamma_{o}}{k_{G}}\right) \left(\frac{n_{G}}{n_{o}}\right)^{2p} - \left(\frac{k_{G}}{\gamma_{o}}\right) \left(\frac{n_{o}}{n_{G}}\right)^{2p}\right] \Bigg\rbrace = 0.
\end{align}
We obtain separate equations for the $TE$ and $TM$ solutions, one when $p=0$ and the other when $p=1$. These dispersion equations can be solved numerically by means of iterative root-finding methods, such as those described in \cite{56,57}. This means, that the value of $\beta$ for the $TE/TM$ supermode at a given frequency $\omega$, must be found through a numerical iteration. The numerical solutions are valid for both, weakly guiding two-mode waveguides and directional couplers, as shown in the structure of Fig.\ref{f1}, and for strongly guiding two-mode structures, which occur when the gap $d$ is equal to zero. In order to test the above situations, we reproduced two examples from previous studies by Nolting et al \cite{53} and Moiseyev et al \cite{46}, using Eq.\eqref{5} with system parameters determined on the values reported in those references, but, with a notorious difference between weakly guiding and strongly guiding two-mode waveguides, where the gain-loss parameter definition in the latter one has been considered as $n_{I}=\frac{\lambda}{4 \pi} \alpha$ instead of $n_{I}=\frac{\lambda}{2 \pi} \alpha$. The propagation constants of the fundamental and first higher-order supermodes, denoted $TE_{0}/TM_{0}$ and $TE_{1}/TM_{1}$, respectively, are depicted in Figure \ref{f2}, with each configuration operating at a wavelength of $2\pi/k = 1.55 \mu\mathrm{m}$. The graph plots these propagation constants against the gain-loss parameter $\alpha$.\\
Fig.\ref{f2} (a) and (b) show how the propagation constants for the $TE_{0}$ and $TE_{1}$ supermodes, blue and purple lines, remain completely different until a critical value of $\alpha$. Notably, as we increase the loss and gain of the material, the two $TE$ supermodes come near each other until the modes merge forming an EP, which emerges at $\alpha \approx 5230\, \mathrm{cm}^{-1}$ for the strongly guiding coupling scenario, and $\alpha \approx 8.398\, \mathrm{cm}^{-1}$ for the weakly guiding coupling case. As it is evident from both waveguide configurations, below the $\mathcal{PT}$-symmetric exceptional point, the propagation constants of $TE_{0}$ and $TE_{1}$ are purely real and this region is referred to as the unbroken $\mathcal{PT}$ symmetry domain. At the threshold point, the real propagation constants of both $TE$ modes become equal, and beyond it, the non-Hermitian system undergoes spontaneous $\mathcal{PT}$ symmetry breaking, where the $TE$ supermodes cease to be real and become complex conjugate to each other; in this case, both waveguides structures support one gain guiding and loss guiding mode. Although we only show the positive real part of $\beta^2-k^2 n^2_{o}$, we stress that the propagation constants also branches for its imaginary values, $\Im\left(\beta^2-k^2 n^2_{o}\right)$, in the EP and after of it, because of their reciprocity. Same kind of bifurcations, as those described for the $TE_{0}$ and $TE_{1}$ solutions, take place for the pair of $TM_{0}$ and $TM_{1}$ supermodes (blue and purple circles), where the EP occurs at $\alpha \approx 5652 \, \mathrm{cm}^{-1}$ for the strongly guiding coupling case, whereas for the weakly guiding coupling is given at $\alpha \approx 8.442 \, \mathrm{cm}^{-1}$. 
Additionally, it can be demonstrated that both, the $TE$ and $TM$ polarizations, exhibit a symmetric distribution of fields after the EP, because each of the two modes is localized in one of the waveguides, i.e, the amplified supermode is mainly localized in the amplifying waveguide in region II, and symmetrically, the attenuated supermode is mainly confined to the lossy waveguide situated in region IV. The graphs in Fig.\ref{f3} display such characteristic features, where it is evident that one of the modes is attenuated, while the other is amplified as they propagate along the slab waveguide structure. A remarkable point is that at the EP, the field distributions, for both $TE$ as well as for $TM$ polarizations, become identical while still conforming to the orthogonality condition \cite{32}. The transversal distributions of electric and magnetic fields, of both structures, have been calculated numerically from Eq.\eqref{2}, by using the propagation constants results of its supermodes from Fig.\ref{f2}; details about the calculation of the unknown coefficients of the fields are presented in Appendix C. Eventually, one can also calculate, and visualize, the total power distribution for the stationary eigenmodes, $\abs{\Phi_{y}(x,z,t)}^2=\abs{\frac{1}{\sqrt{2}}  \bigg( \tilde{\phi_{0}}(x) e^{-i\beta_{TE_{0}} z} + \tilde{\phi_{1}}(x) e^{-i\beta_{TE_{1}} z}  \bigg) e^{i \omega t} }^2$, with $\tilde{\phi_{j}}(x)=\frac{\phi_{j}(x)}{N_{j}}$, being $N_{j}$ the $C$ norm \cite{55} defined as $N_{j}=\int_{-\infty}^{\infty} \phi^{2}_{j}(x) dx$,  for the $\mathcal{PT}$-symmetric waveguide systems in the vicinity of the EP. It should be noted that when dealing with propagating modes, the boundary conditions require that $\phi_{y}(x)$ goes to zero as $x$ approaches positive or negative infinity; the power distribution is displayed in Figs.\ref{f4} (a) and (b) for the two guided $TE$ supermodes of each waveguide configuration. There, it is shown that the total field travels simultaneously from the gain waveguide to the loss waveguide and vice versa, resulting in light oscillations with a beat length of $L=2\pi/\Delta\beta$, where $\Delta\beta$ represents the difference between two modes. By adjusting and increasing the gain-loss parameter $\alpha$, one can directly observe the propagation constant movement towards the exceptional point as $\Delta\beta$ becomes smaller, making it possible to study the phenomenon by varying a single parameter on the $\mathcal{PT}$-symmetric slab system. It is noteworthy that upon reaching the EP, when $\Delta\beta \approx 0$ (the beat length becomes infinite), the power distribution shows a pulsing pattern in both waveguides simultaneously, rather than oscillating between them.

\section{Group velocity in a balanced gain-and-loss slab waveguide}
\begin{figure}[H]
    \centering
    	\subfloat[]
	{\includegraphics[width=0.48\textwidth]{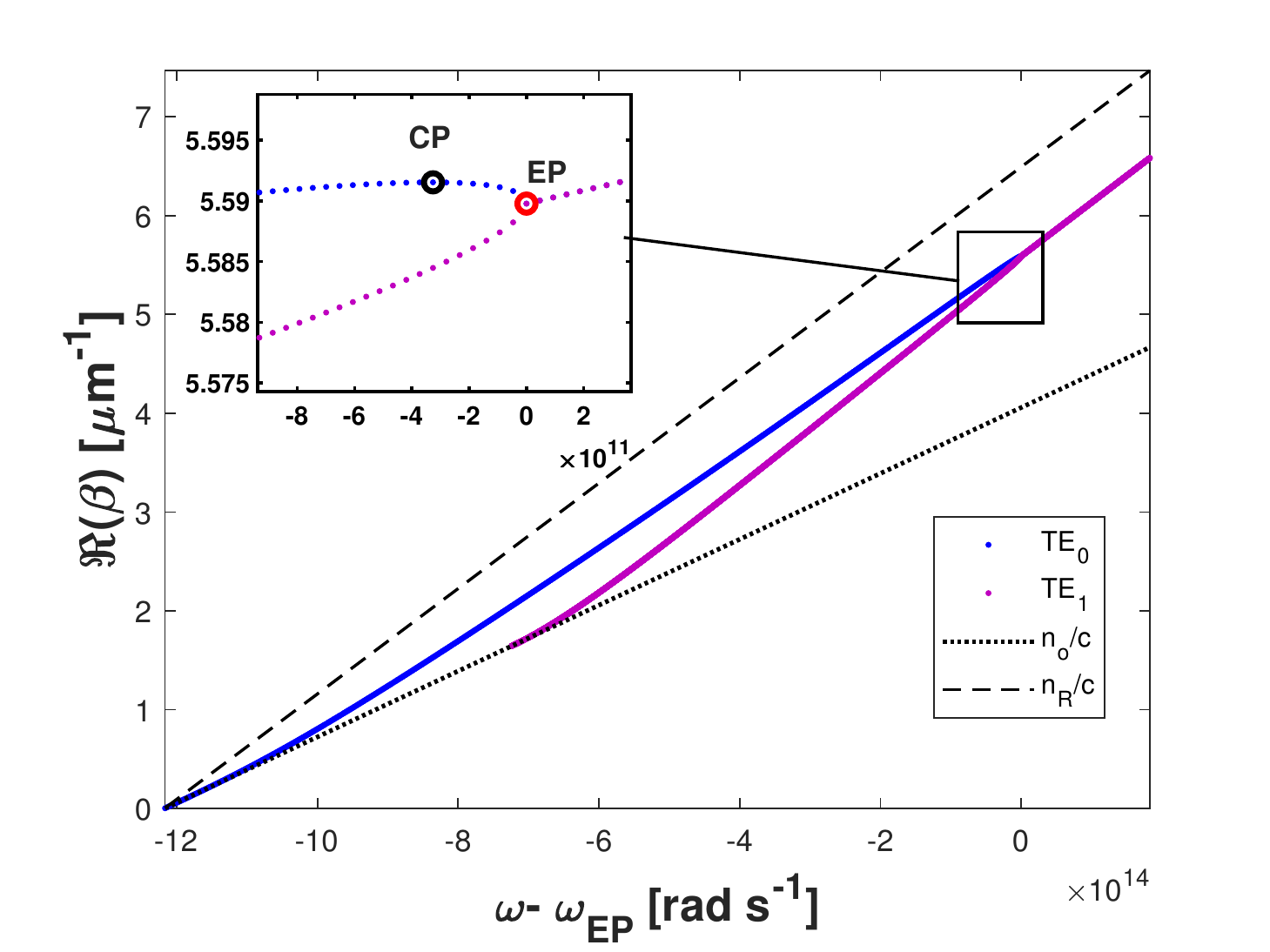}}
		\quad
	\subfloat[]
	{\includegraphics[width=0.48\textwidth]{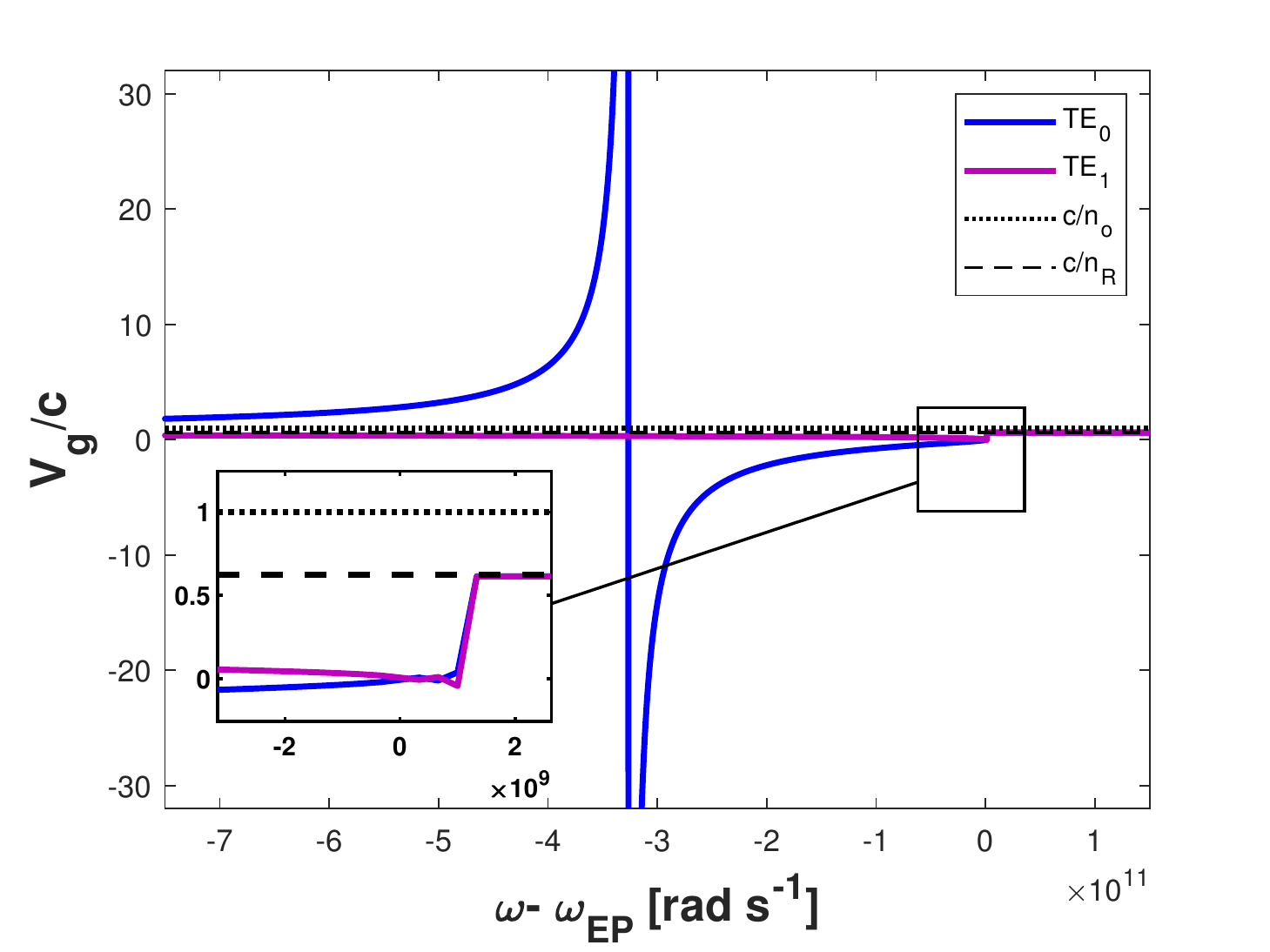}}
		\quad
	\subfloat[]
	{\includegraphics[width=0.48\textwidth]{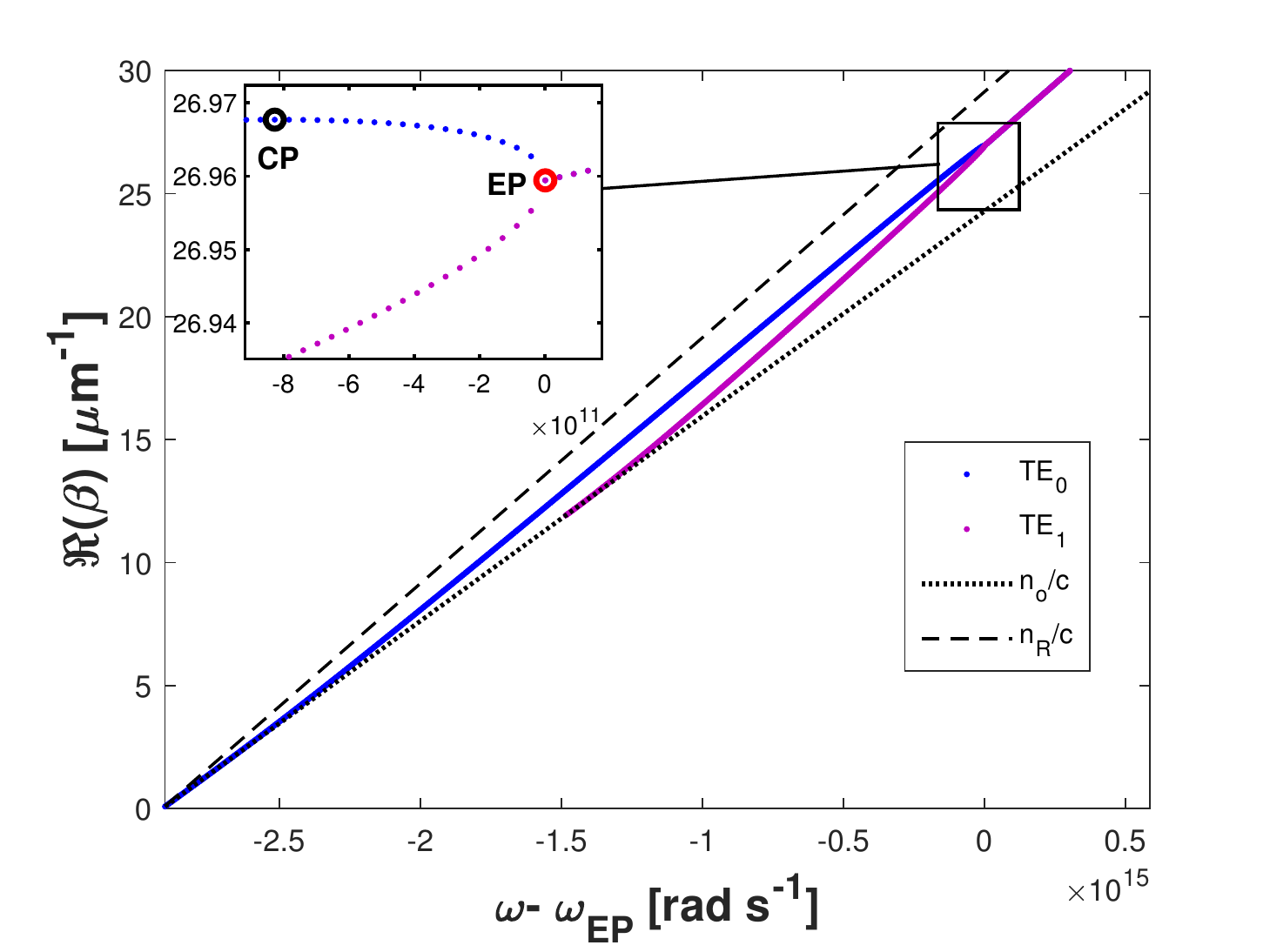}}
		\quad
	\subfloat[]
	{\includegraphics[width=0.48\textwidth]{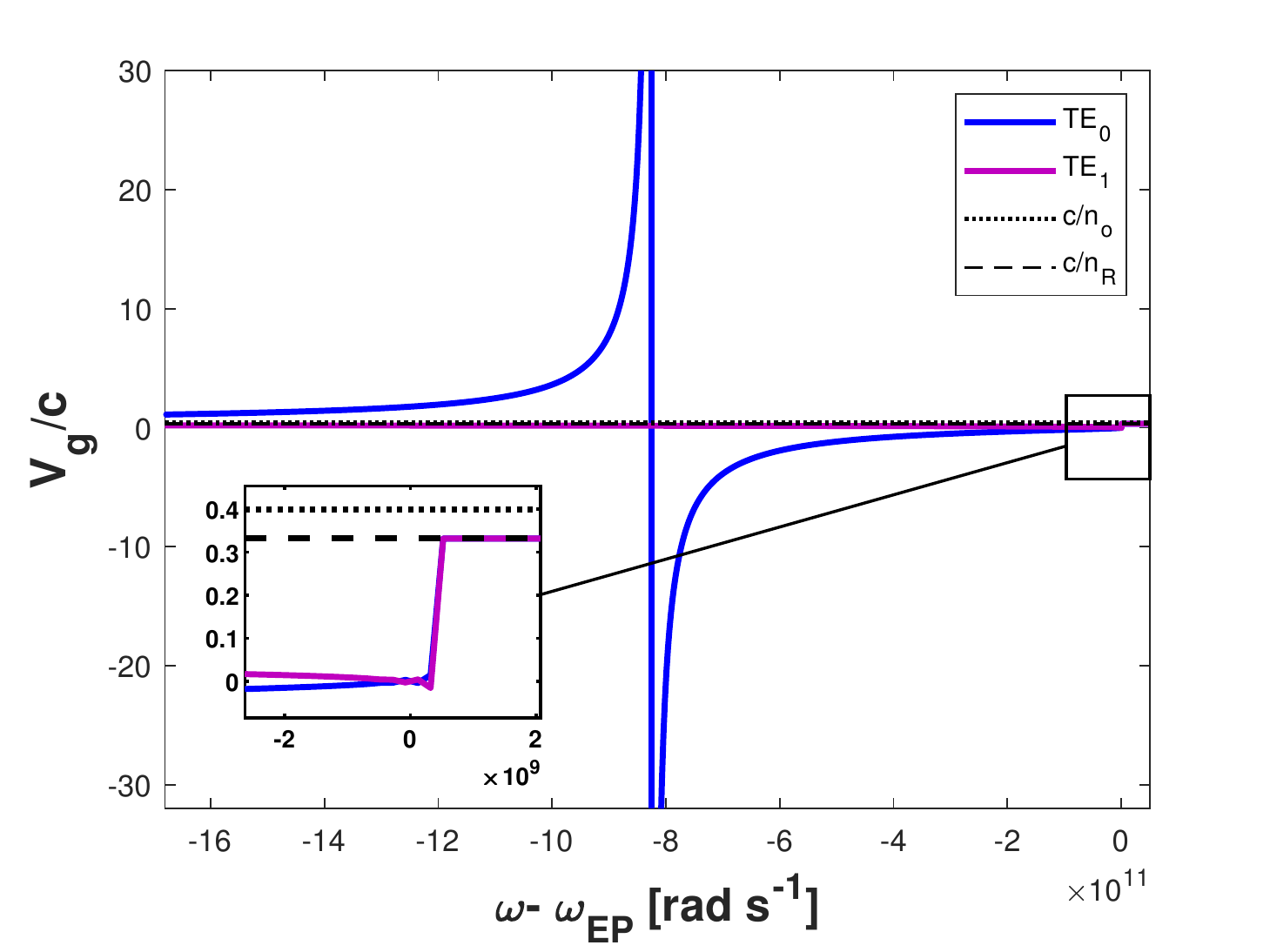}}
	\caption{(a) and (c) show the dispersion diagrams of the real part of the propagation constant as function of frequency $\omega$. These diagrams exhibit the $\beta$ trajectories of the pair of supermodes, the fundamental $TE_{0}$ and the first higher-mode $TE_{1}$, approach to each other until they simultaneously coalesce at EPs, such as marked in red in the zoomed parts in the plots. At these points, both $TE$ propagation constants degenerate to a single real value. (b) and (d) show the group velocities versus the angular frequency $\omega$ for the $TE_{1}$ and $TE_{0}$ supermodes in the non-Hermitian system. Around the CP, the $TE_{0}$ group velocity is speed up, corresponding to a fast light effect. After the CP, the fundamental supermode exhibits a negative group velocity related to backward light, and then switches to subluminal light near the EP. Numerical solutions of (a) and (b) were obtained using the parameters $n_{R}=1.6$, $n_{I}=0.02$, $n_{o}=1$ and $W=d=0.524\, \mu\mathrm{m}$, meanwhile for (c) and (d) were obtained by adjusting the set of values to $n_{R}=3$, $n_{I}=0.05$, $n_{o}=2.5$, $W=0.15\, \mu\mathrm{m}$ and $d=0.1\, \mu\mathrm{m}$. In both cases, a root-finding method was applied in Eq.\eqref{5} to separately calculate 30,000 roots for each $TE$ supermode.}
	\label{f5}
\end{figure}

It should be remarked that the above results are not the only information that one can get from such slab waveguides, especially when $n_{I}$ (in terms of $\alpha$) operates as the only parameter. In particular, the modal analysis applied here, gives us the freedom to kept $n_{I}$ at a fixed value, and get the physical picture of the $\beta-\omega$ dispersion diagram for the lowest $TE$ supermodes, which have their own propagation constants, and of course, that travel with their own velocity. This dispersion diagram offers valuable information, such as the group velocity of each supermode, which can be determined by analyzing the slope of $V_{g}=\left(\frac{d \beta}{d \omega} \right)^{-1}$. To further elucidate this aspect, we are going to consider two specific sets of parameters, whose $n_{I}$ values are within a range of experimentally achievable data \cite{58,59,60}; the choice of the parameters is done to examine closely the behavior of the group velocities for the fundamental mode and the first higher-order mode, over a medium and high-frequency range. For the first set of parameters, we utilize the following values in Eq. \eqref{5}: $n_{R}=1.6$, $n_{I}=0.02$, $n_{o}=1$ and $W=d=0.524 \, \mu \mathrm{m}$; Fig.\ref{f5}(a) illustrates the results of the settings used in this configuration. It can be observed that the actual propagation constants of dispersion for the $TE$ eigenmodes gradually approach one another, ultimately leading to the merging of the two dispersion branches and the formation of an EP at a frequency $\omega_{EP} = 1.217x10^{15}$ rad $\mathrm{s}^{-1}$, which corresponds to a wavelength of $\lambda_{EP}=1.549 \mu \mathrm{m}$. The inset in Fig.\ref{f5} (a) shows a zoomed-in view of the crossing point EP, which is encircled in red; these supermodes exist in the region between lower $n_{o}/c$, and upper cutoff frequencies $n_{R}/c$, denoted by dotted and dashed lines, respectively; for practical purposes, we have omitted $TM$ solutions to easily distinguish the trajectories of the propagation constants of $TE$ supermodes. Once the propagation constants of the $TE_{0}$ and $TE_{1}$ supermodes are determined for each value of $\omega$, the associated group velocity $V_{g_0,g_1}=\left(\frac{d \beta_{1,2}}{d \omega} \right)^{-1}$ of each supermode, as  function of angular frequency, can be easily simulated by using the 5-point finite difference approximation\cite{61,62,63}. In order to see a substantial effect, the numerical solutions of the group velocity of both supermodes are calculated in the same frequency window as the one in the zoomed region of Fig.\ref{f5}(a); these solutions are graphically illustrated in Fig.\ref{f5}(b). Interestingly, we find that the group velocity for the $TE_{0}$ supermode starts with a large and positive anomalous dispersion near the frequency $\omega_{CP}=1.2166x10^{15}$ rad $\mathrm{s}^{-1}$, and which we will refer to as the critical frequency or the critical point (CP); there, the exotic light transition occurs. However, this superluminal propagation tendency changes abruptly after the CP, where the $TE_{0}$ group velocity experiences a negative and large group velocity, $V_{g}<c$, overcoming the minimum group velocity corresponding to the bottom line, $V_{gmin}=c/n_{R}$. This behavior gives rise to a backward subliminal light \cite{64,65,66}, that near to the EP switches to subluminal light ($0< V_{g} \ll c$). Such abrupt shifts in light behavior occur between the CP and the EP, inside of the unbroken $\mathcal{PT}$-symmetric domain; the CP has been identified and marked by a black circle in the zoomed frequency window of Fig.\ref{f5}(a). In contrast to the $TE_{0}$, the group velocity for $TE_{1}$ exhibits a slow light effect at all $\mathcal{PT}$-symmetric domains during its propagation.\\
The same behavior of the group velocities occurs when one chooses another set of values in the slab waveguide system. The dispersion diagram shown in Fig.\ref{f5}(c) supports these predictions, and it has been obtained replacing in Eq.\eqref{5} the parameters $n_{R}=3$, $n_{I}=0.05$, $n_{o}=2.5$, $W=0.15 \, \mu\mathrm{m}$ and $d=0.1\, \mu\mathrm{m}$. As shown in Fig.\ref{f5}(d), the group velocity of the fundamental supermode again displays a fast light character near to the CP at $\omega_{CP}=2.9143x10^{15}$ rad $\mathrm{s}^{-1}$, and the backward light propagation occurs in the region between the CP and the EP. In contrast to the first waveguide setup, this second set in the waveguide hosts an EP localized in a high-frequency  $\omega_{EP}=2.9151x10^{15}$ rad $\mathrm{s}^{-1}$, corresponding to $\lambda_{EP}=0.6466\, \mu\mathrm{m}$; moreover, our numerical solutions are in excellent agreement with the ones reported by Lingxuan\cite{49}; however, our method allows to get solutions in the $\mathcal{PT}$-unbroken and broken regions, and likewise, without neglecting the material and geometry parameters; for instance, the background medium wherein the two slab guiding layers are embedded. In fact, one can also replicate Lingxuan's original findings by substituting the set of the parameters given in reference \cite{49} into Eq.\eqref{5}, but taking into account that the background index is equal to 1. On the other hand, the results of the numerical simulations of Fig.\ref{f5}, also confirm that a $\mathcal{PT}$-symmetric optical waveguide, working near the EP, could be used to slow down a pulse of light, such as have been recently reported by Moiseyev et al\cite{55}; this can be judged quickly in the inset zoomed region from Figs.\ref{f5} (b) and (d). 
The numerical results suggest that the group velocities of both supermodes go to zero and then start to increase monotonically until both group velocities are equal, leading the group velocity of the light pulse to become constant and propagates within the cutoff minimum group velocity $V_{gmin}$. It should be noted that in contrast to Moiseyev's results, ours are not restricted to assuming that the EP is outside of the operation protocol, unlike \cite{55}, where it has been excluded to avoid instabilities and get the dispersion curve of the propagation constants. Further, in all our calculations, we keep $n_{I}$ fixed instead of changing at each frequency due to $n_{I}=\frac{\alpha}{k}$, being a contrasting character between our work and those presented in \cite{55}; albeit, if we take the latter consideration in our numerical simulations, we will obtain the same behavior for the group velocities. It is worth mentioning, that a typical limitation to practical applications and for an experimental setup of the above waveguide model is that it requires a fine adjustment of the system parameters, which is quite difficult to reach, specially for the attainability of slow and fast light effects near to the EP.\\
\begin{figure}[H]
    \centering
    	\subfloat[]
	{\includegraphics[width=0.48\textwidth]{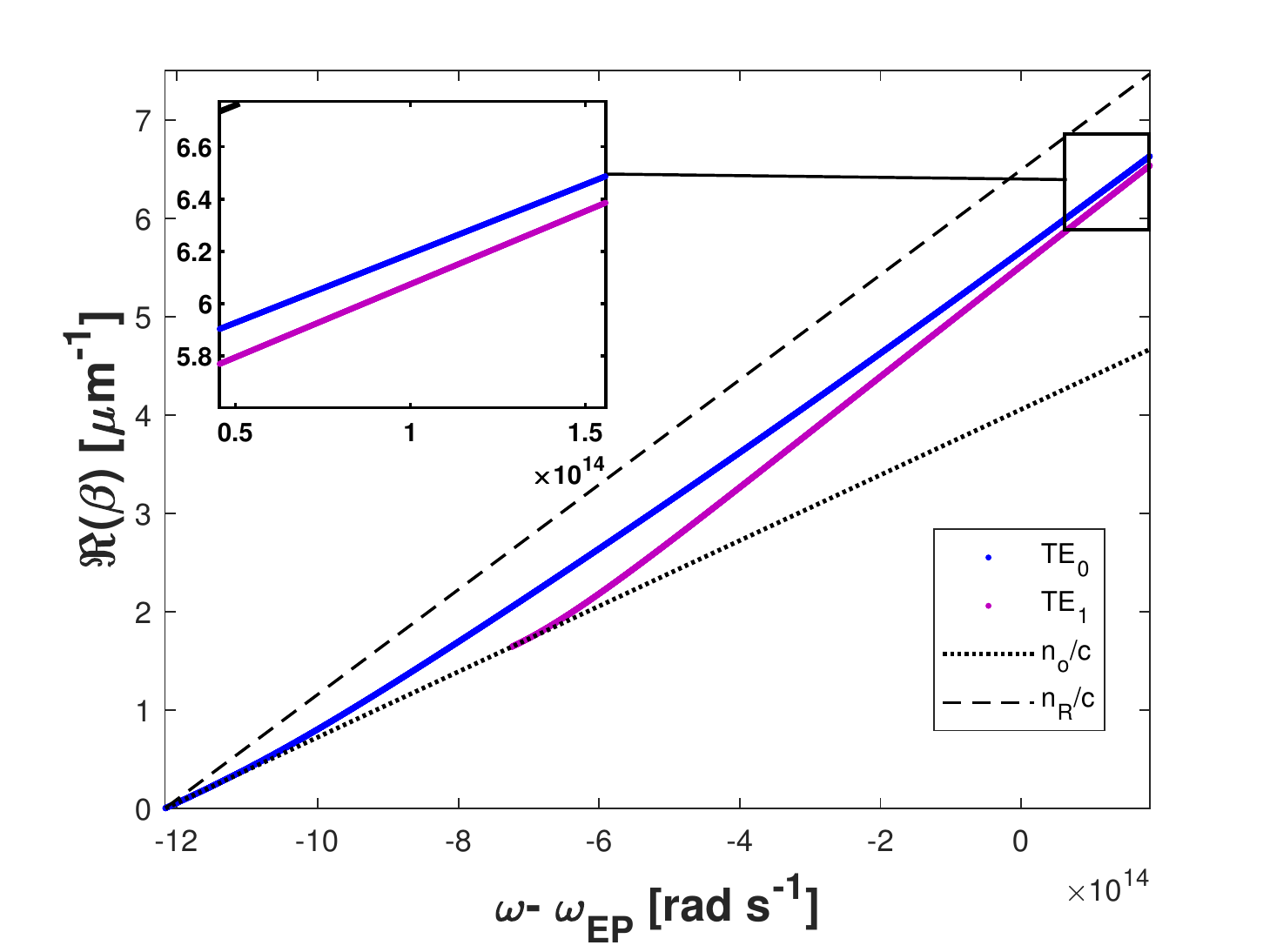}}
		\quad
	\subfloat[]
	{\includegraphics[width=0.48\textwidth]{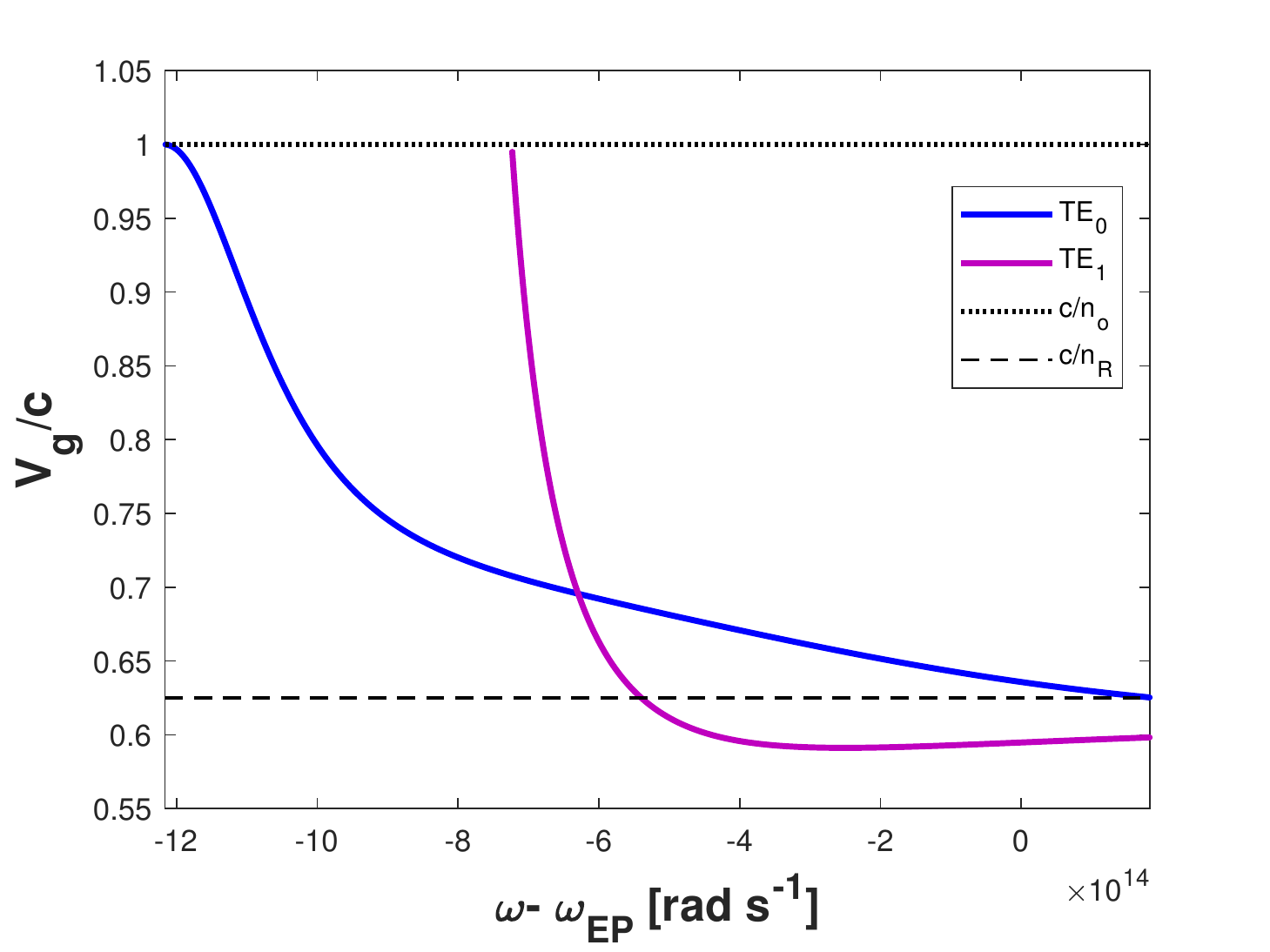}}
		\quad
	\subfloat[]
	{\includegraphics[width=0.48\textwidth]{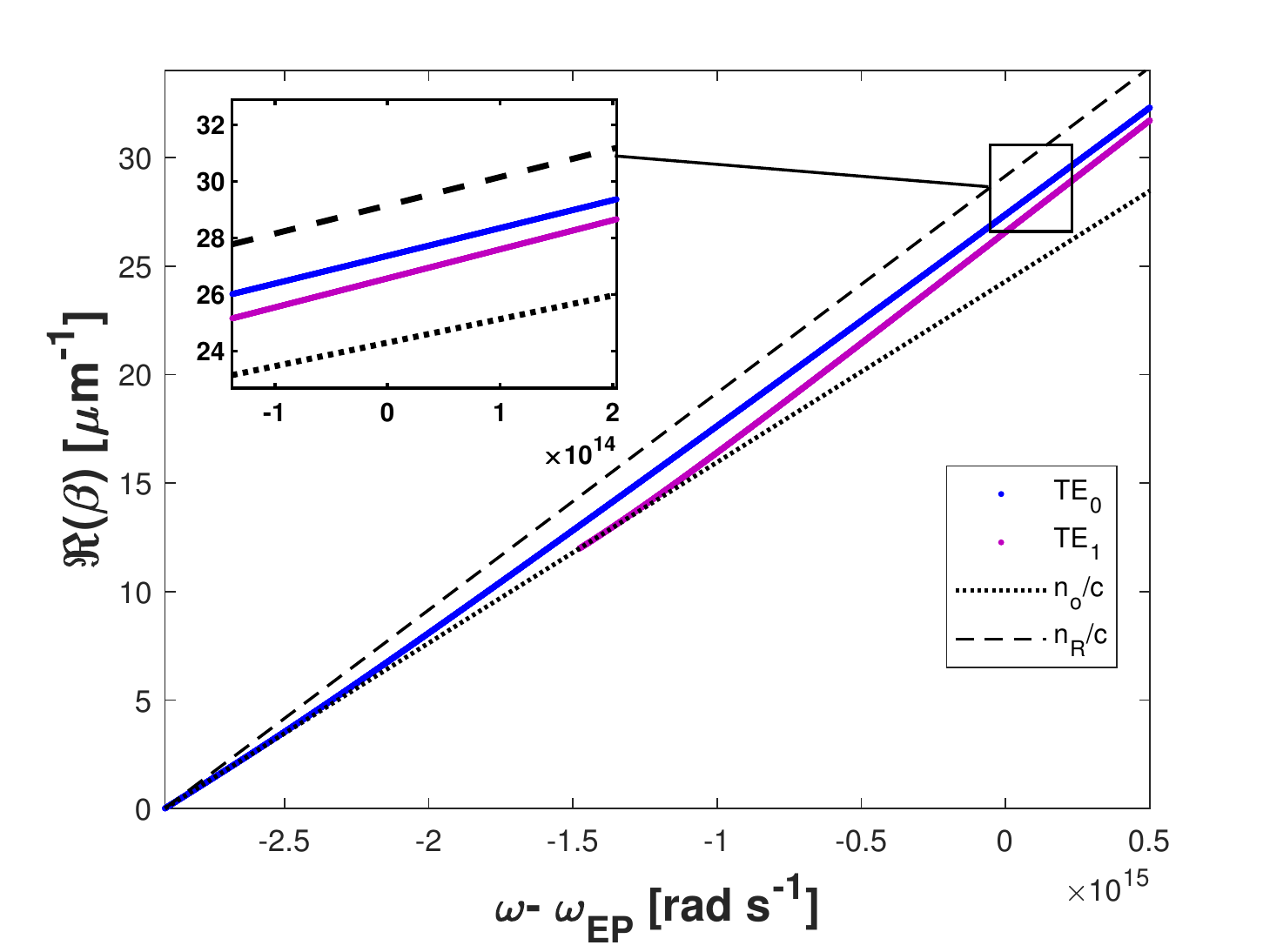}}
		\quad
	\subfloat[]
	{\includegraphics[width=0.48\textwidth]{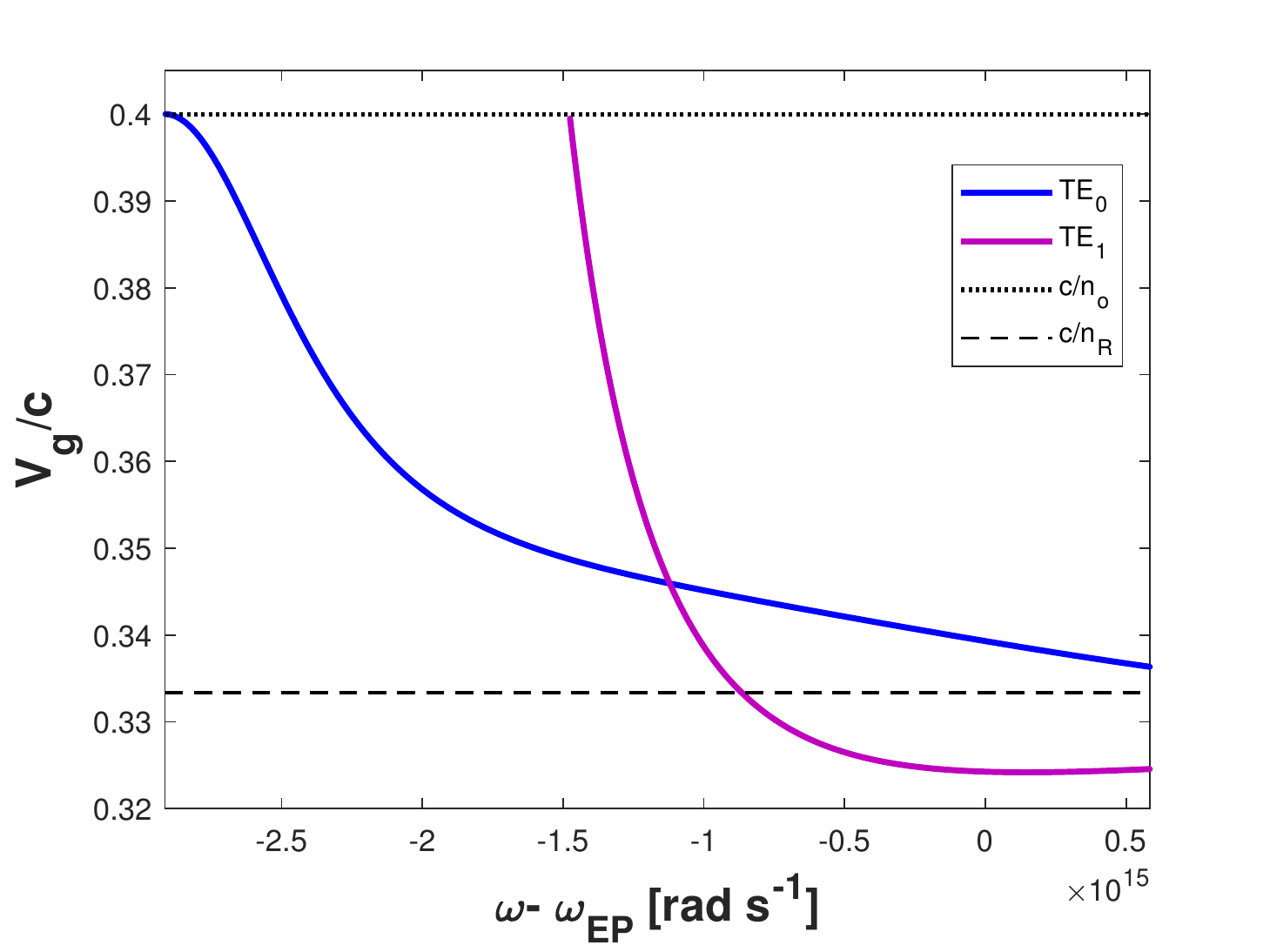}}
	\caption{(a) and (c) show the dispersion diagrams of the real part of the propagation constant $\beta$ as a function of frequency $\omega$. These graphs outline the dispersion curve behavior when the gain-loss parameter is not considered in the two-slab waveguide system, i.e., $\alpha=0$. As intuitively expected, the $TE$ eigenmodes coalescence does not take place, such as it is observed in the zoomed-in view of (a) and (c). Here, the eigenmodes become symmetric and antisymmetric supermodes. Since the waveguide system is now Hermitian, the EP can no longer be formed; this leads to the group velocities following the usual shape and slope; (b) and (d) illustrate this fact.}
	\label{f6}
\end{figure}
As a sanity check, we also provide a comparison of the dispersion diagrams and group velocities when the gain-loss parameter, $\alpha$, is equal to zero. Figs.\ref{f6} (a) and (c) show that the pairs of $TE$ eigenmodes in the two-slab model never coalesce; this feature can be clearly appreciated from the inset zoomed-in view of both graphs. In this case, the Hermitian waveguide system possesses a pair of symmetric and antisymmetric modes; then, the zoomed regions from Figs.\ref{f6} (b) and (d) show that the group velocity for $TE_{0}$ is near to the cut-off of the maximum group velocity $V_{gmax}$ and keeps inside of the slope of the minimum group velocity $V_{gmin}$, whereas, for the $TE_{1}$, the group velocity falls outside the range of $V_{gmin}$. Despite the above, the group velocities do not appear to indicate any anomalous deviation between the slope, $c/n_{o}$, for the background material and for the core slab, $c/n_{R}$; therefore, their group velocities follow a typical behavior similar to those observed in uncouple waveguides.\\
It is important to point out that the non-Hermitian subliminal light at the EP and the superluminal group velocity near it, are unconventional interference phenomena that cannot be achieved within a single waveguide. In fact, it is possible to think that the group velocity loses its usual physical meaning between the CP and the EP due to the two aforementioned unconventional effects. Specifically, the group velocity in absorbing and active media is defined similarly to the Hermitian systems, and this could lead to the physical interpretation of the group velocity remaining unclear about how the energy transports in non-Hermitian systems; however, authors in \cite{67} have stated a theoretical proposition to clarify the energy interpretation of the dynamics of optical waves in coupled $\mathcal{PT}$-symmetric waveguides, where they state that one must relinquish the notion that the electromagnetic fields in nature are purely real rather than complex identities. The same applies to the energy flux, defined by the Poynting vector, for dealing with non-Hermitian optics; for instance, they have assumed the conundrum of the zero group velocity at EP is a consequence of the instantaneous complex Poynting vector which is also equal to zero, whereas the superluminal group velocity relies on the complex instantaneous local energy density, which could be negative to explain the fast light effect. Even though our results match also those reported in ref \cite{67}, it is important to stress the significance of thoroughly examining the definition of group velocity, when dealing with non-Hermitian optical systems. In order to gain a more comprehensive understanding of the complex group velocity, a more comprehensive reformulation of the definition is necessary. This updated definition could provide not only information regarding energy transport but also about energy dissipation in the medium. We believe that establishing a connection between group velocity in both Hermitian and non-Hermitian cases is crucial to compare the theoretical framework of all reported results near the EP. In particular, a successful observation of stopped light at the EP, and fast light effect in optical experiments, could lend support to previously reported works, including our own.\\
A potential future application for superluminal light could be in a hypercomputation system\cite{68,69,70}, allowing for infinite computation steps to be performed within a finite amount of time, compared to systems based on subluminal photons, which would require an infinite amount of time to complete infinite computation steps. Utilizing superluminal photons for computation could result in energy costs of less than those performed by conventional silicon processors. Furthermore, a significant aspect is that the sudden shift from rapid forward to backward light propagation at the CP has the potential to be utilized for the detection of much smaller changes in the environment, through the modification in the spacing of the waveguides. In addition, it is conceivable that the $\mathcal{PT}$-symmetric slab system can be connected to an ultra-weakly hermitian environment, which would prohibit heat exchange between them. Alternatively, the arrangement would exclusively allow the flow of quantum information in and out. This setup would facilitate the exploration of the phenomenon of critical slowing down of decoherence; in fact, we envision extending our work into the heart of the quantum information revolution. We believe that by combining a cross-cavity with a coupled $\mathcal{PT}$-symmetric slab system, a novel quantum information processing approach can be developed. A cross-cavity is an optical cavity that allows light to be coupled between two perpendicular directions. By introducing the cross-cavity between the two waveguides and coupling the light between the two systems, it may be possible to create a reliable quantum configuration for computing purposes. This can be achieved by using the light from the $\mathcal{PT}$-symmetric slab system to perform quantum logic gates, where the light interacts with the qubits in the cross-cavity. The qubits can be manipulated using the quantum logic gates and the light that passes through the cross-cavity can be coupled back into the waveguides. By interfering with the light from the other waveguide, the cross-cavity could enable parallel quantum computations.

\section{Asymmetric slab thickness case $W_{a} \neq W_{b}$ }
Up until now, we have assumed that the coupled slab waveguide structure has a proportional and equal amount of loss and gain and that the thickness of the two slabs is the same. By satisfying these considerations, the system becomes perfectly $\mathcal{PT}$ symmetric. We have calculated the dispersion diagram and group velocity for the first two supermodes based on these assumptions. However, if the system is imbalanced, some unexpected situations may arise. Specifically, if one waveguide's thickness is slightly different from the other, it can cause an overall imbalance in the slab waveguide system.
\begin{figure}[H]
    \centering
    	\subfloat[]
	{\includegraphics[width=0.48\textwidth]{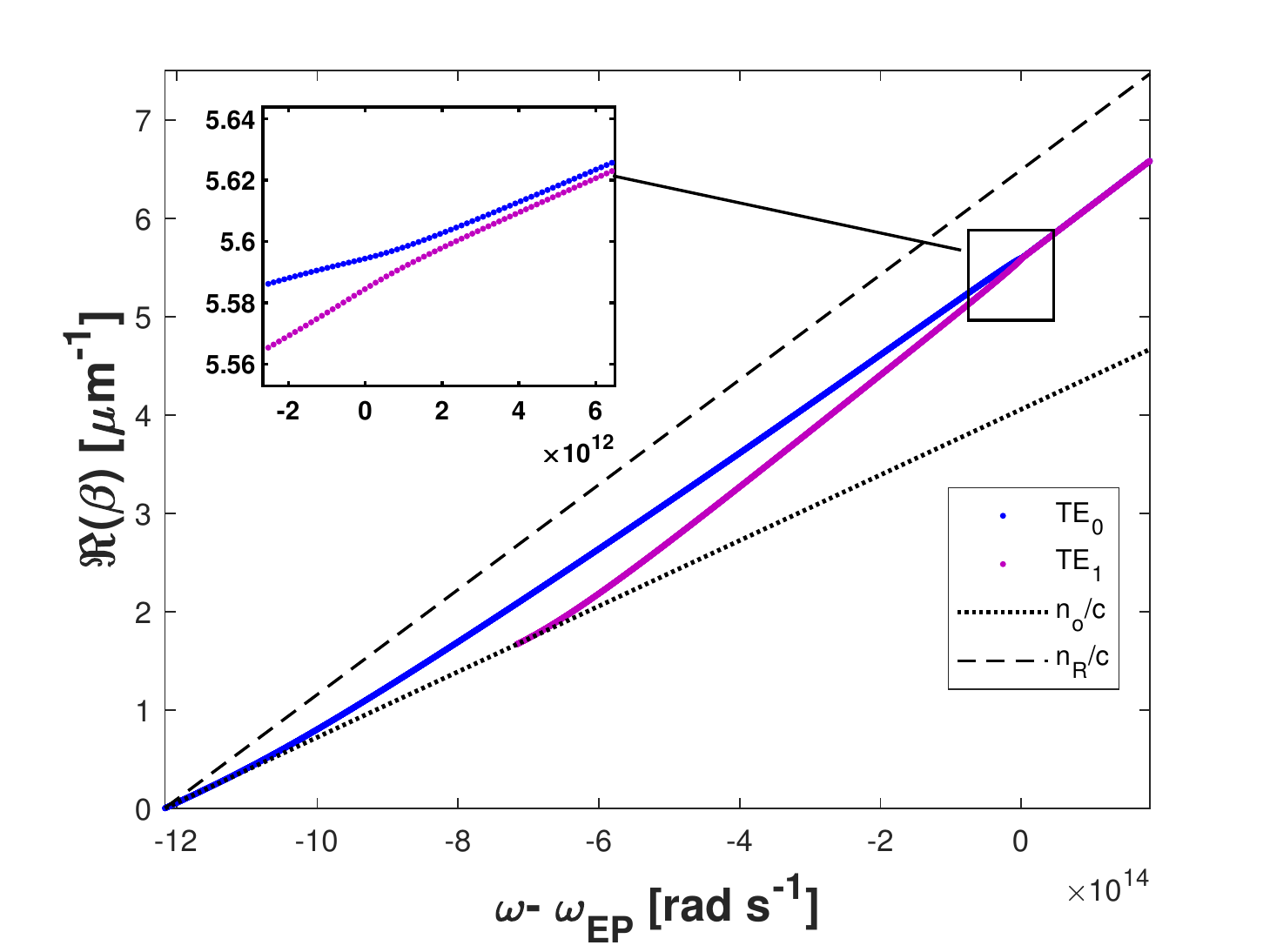}}
		\quad
	\subfloat[]
	{\includegraphics[width=0.48\textwidth]{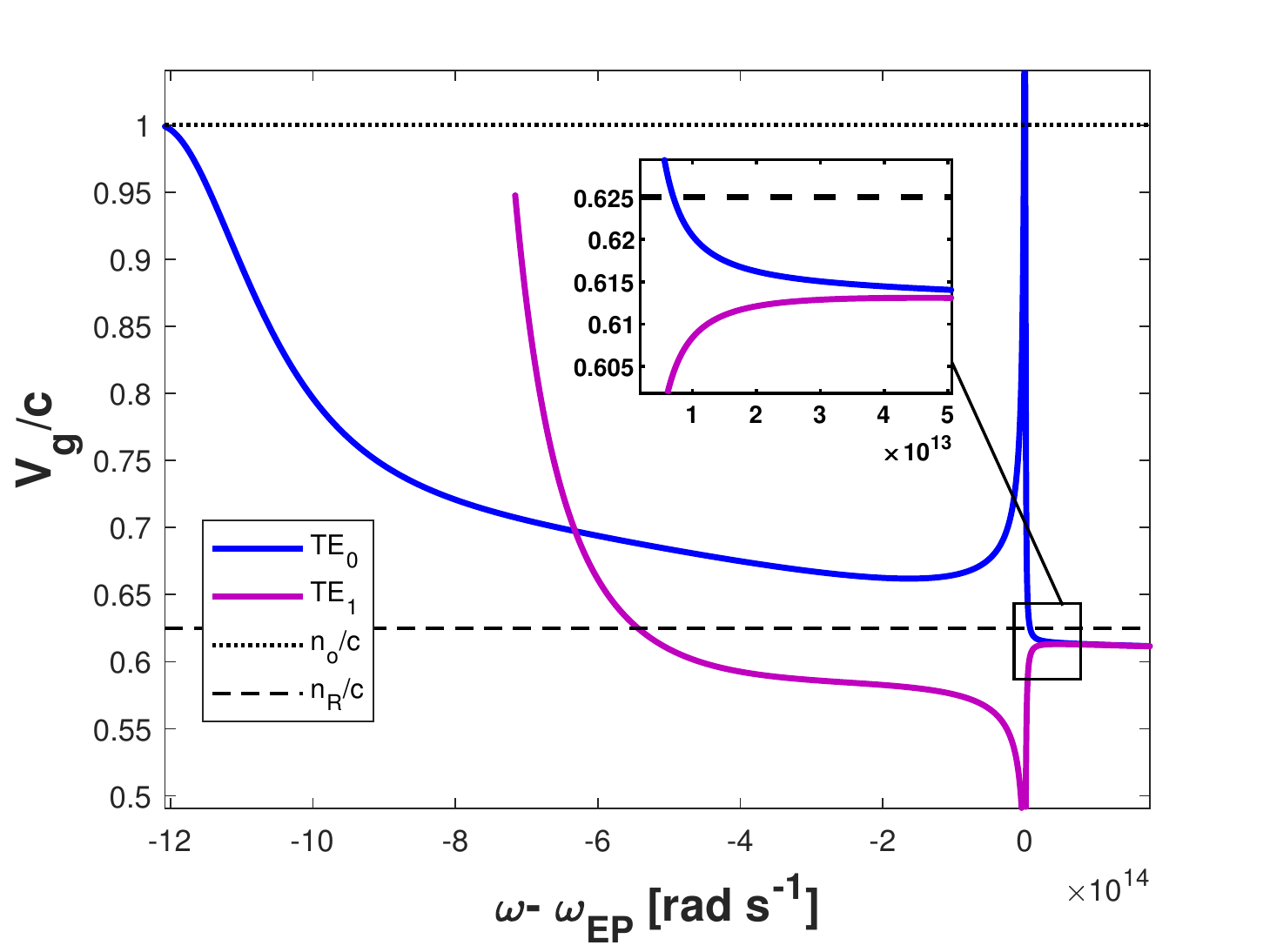}}
		\quad
	\subfloat[]
	{\includegraphics[width=0.48\textwidth]{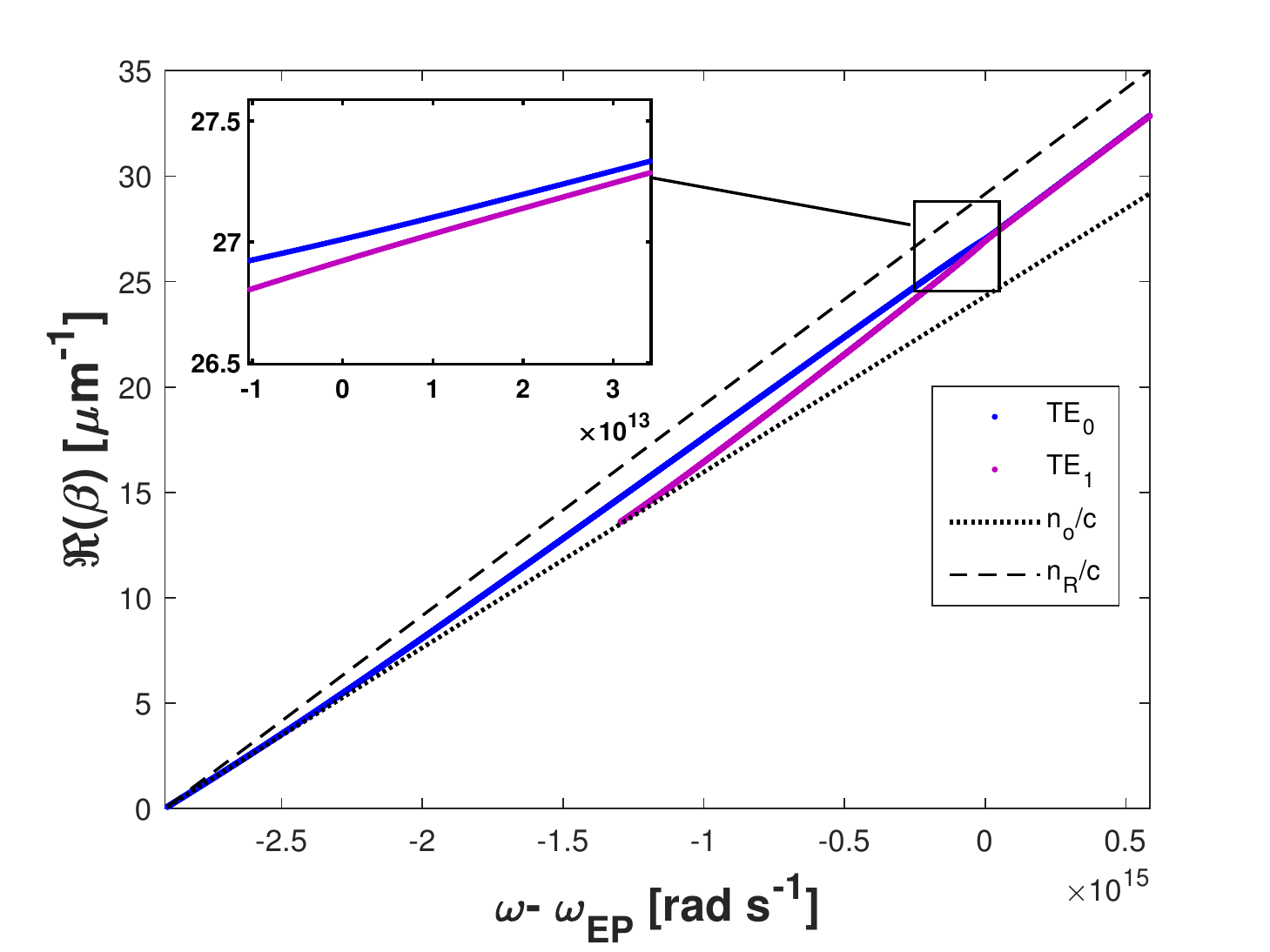}}
		\quad
	\subfloat[]
	{\includegraphics[width=0.48\textwidth]{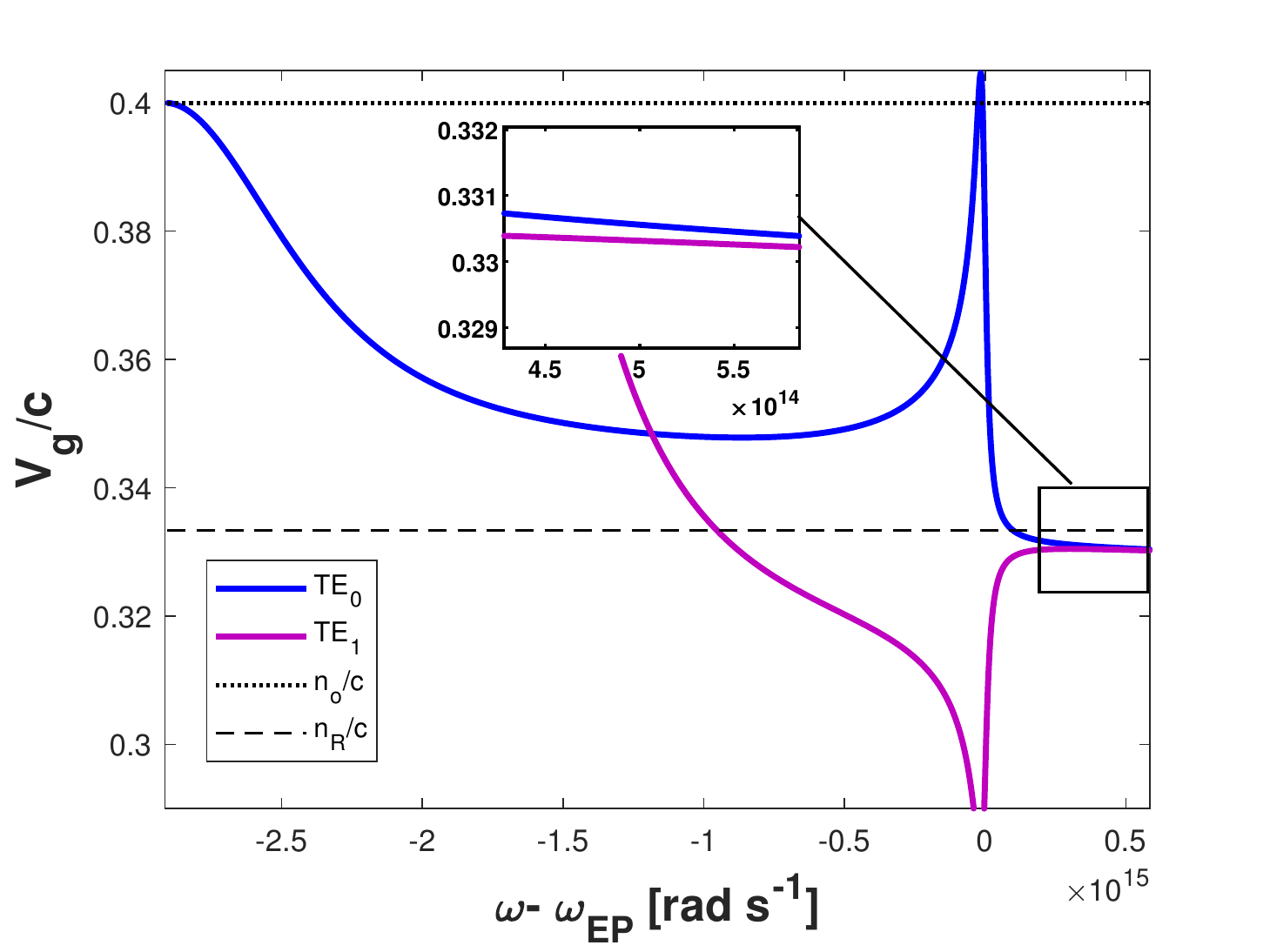}}
	\caption{(a) and (b) show the real part of the trajectories of the propagation constants as function angular frequency $\omega$. It is shown the sensitivity of the non-Hermitian system due to the geometric mismatching between the two slabs'thickness; this detuning avoids that the propagation constants of the $TE_{0}$ and $TE_{1}$ supermodes lead to the formation of the EP. (c) and (b) a slight imbalance in the thickness of the two slabs in the waveguide system results in the evident persistence of both propagations behavior.  The numerical solutions of (a) and (b) were obtained using the parameters $n_{R}=1.6$, $n_{I}=0.02$, $n_{o}=1$, $W_{a}= 0.524 \, \mu\mathrm{m}$, $W_{b}= 0.5237\, \mu\mathrm{m}$ and $d=0.524\, \mu\mathrm{m}$, meanwhile for (c) and (d) were obtained by adjusting to the set of values $n_{R}=3$, $n_{I}=0.05$, $n_{o}=2.5$,  $W_{a}= 0.15 \, \mu\mathrm{m}$, $W_{b}= 0.151\, \mu\mathrm{m}$ and $d=0.1 \, \mu\mathrm{m}$. }
\label{f7}
 \end{figure}
 
\begin{figure}[H]
    \centering
    	\subfloat[]
	{\includegraphics[width=0.48\textwidth]{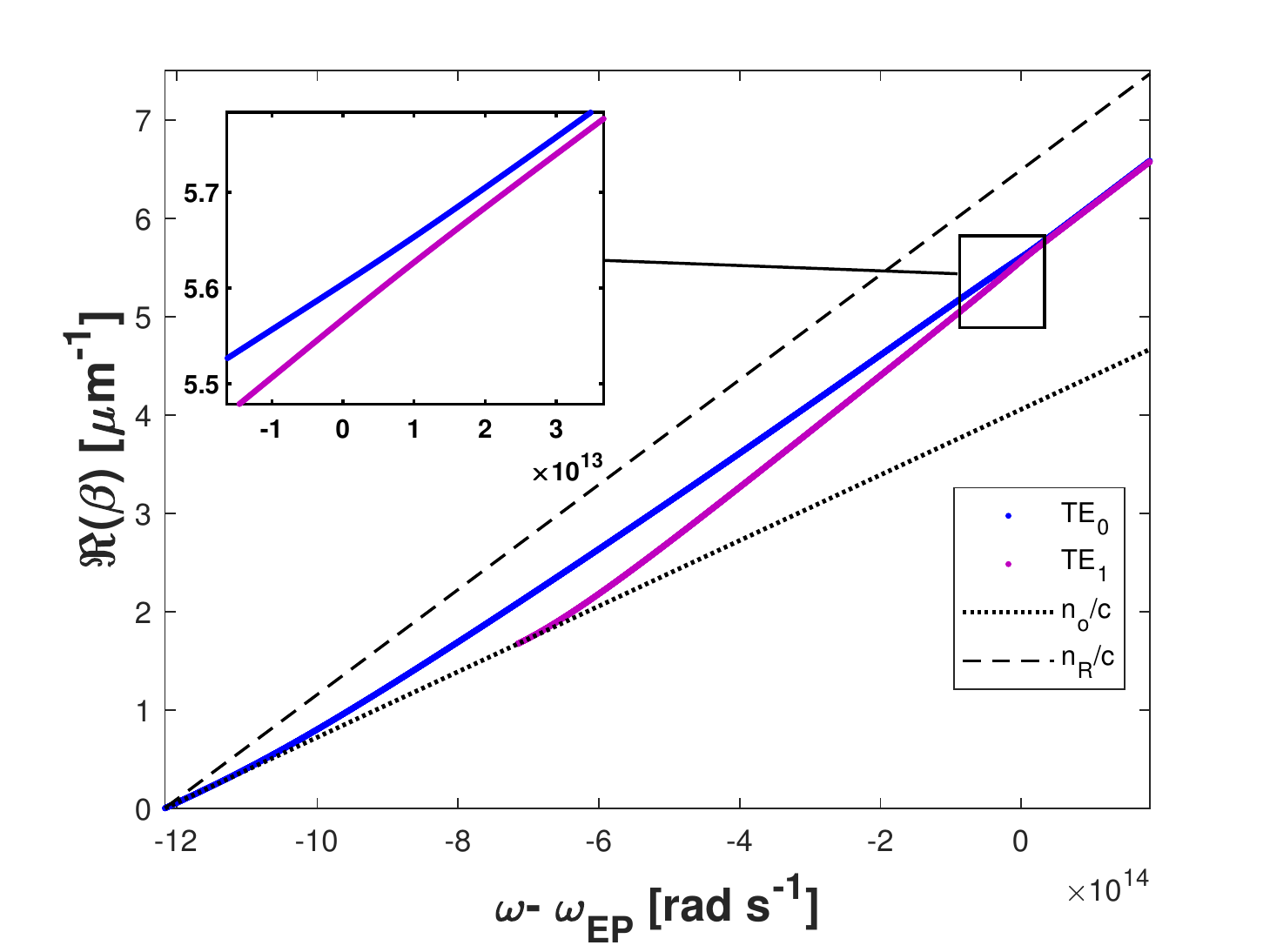}}
		\quad
	\subfloat[]
	{\includegraphics[width=0.48\textwidth]{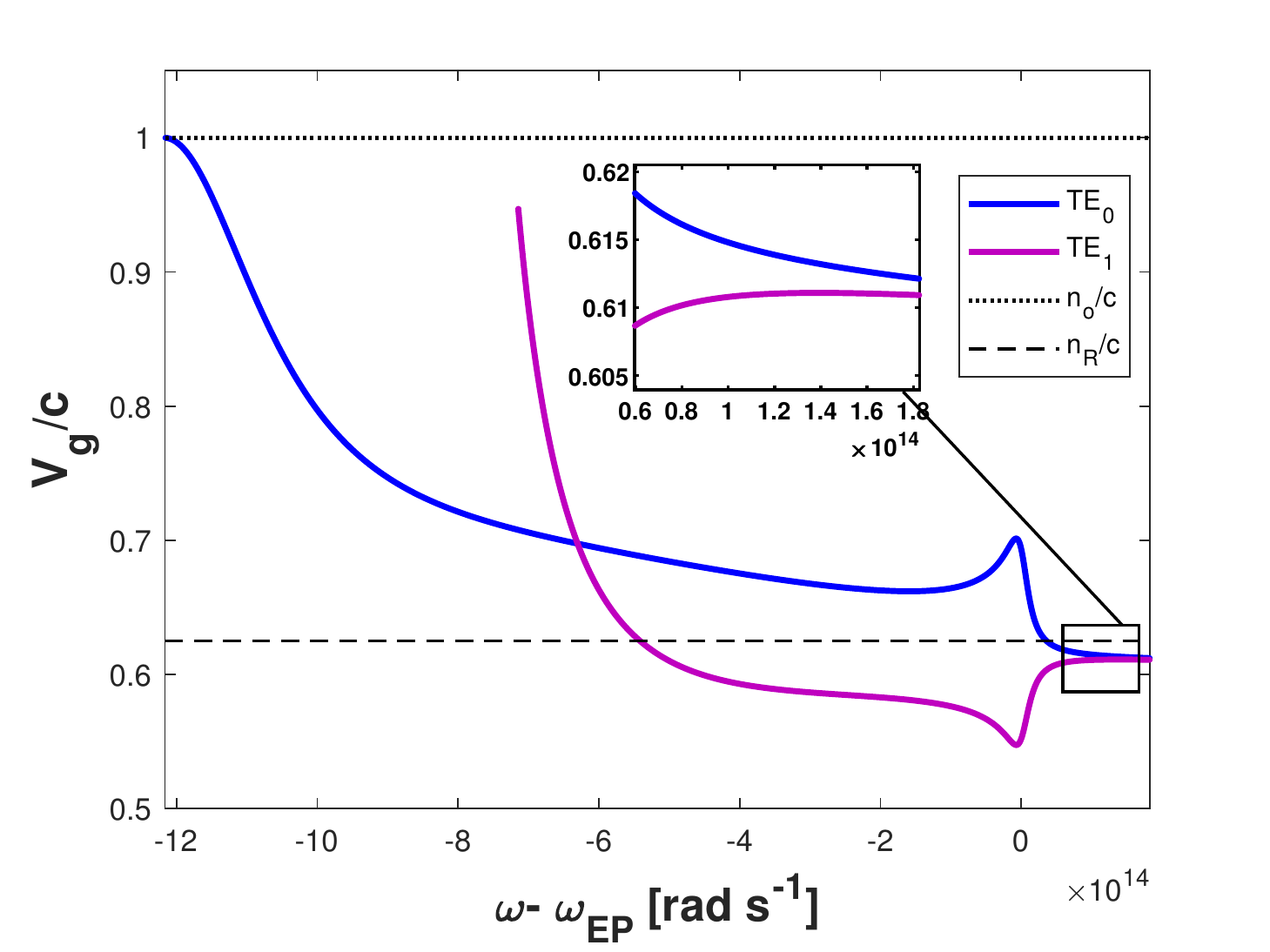}}
		\quad
	\subfloat[]
	{\includegraphics[width=0.48\textwidth]{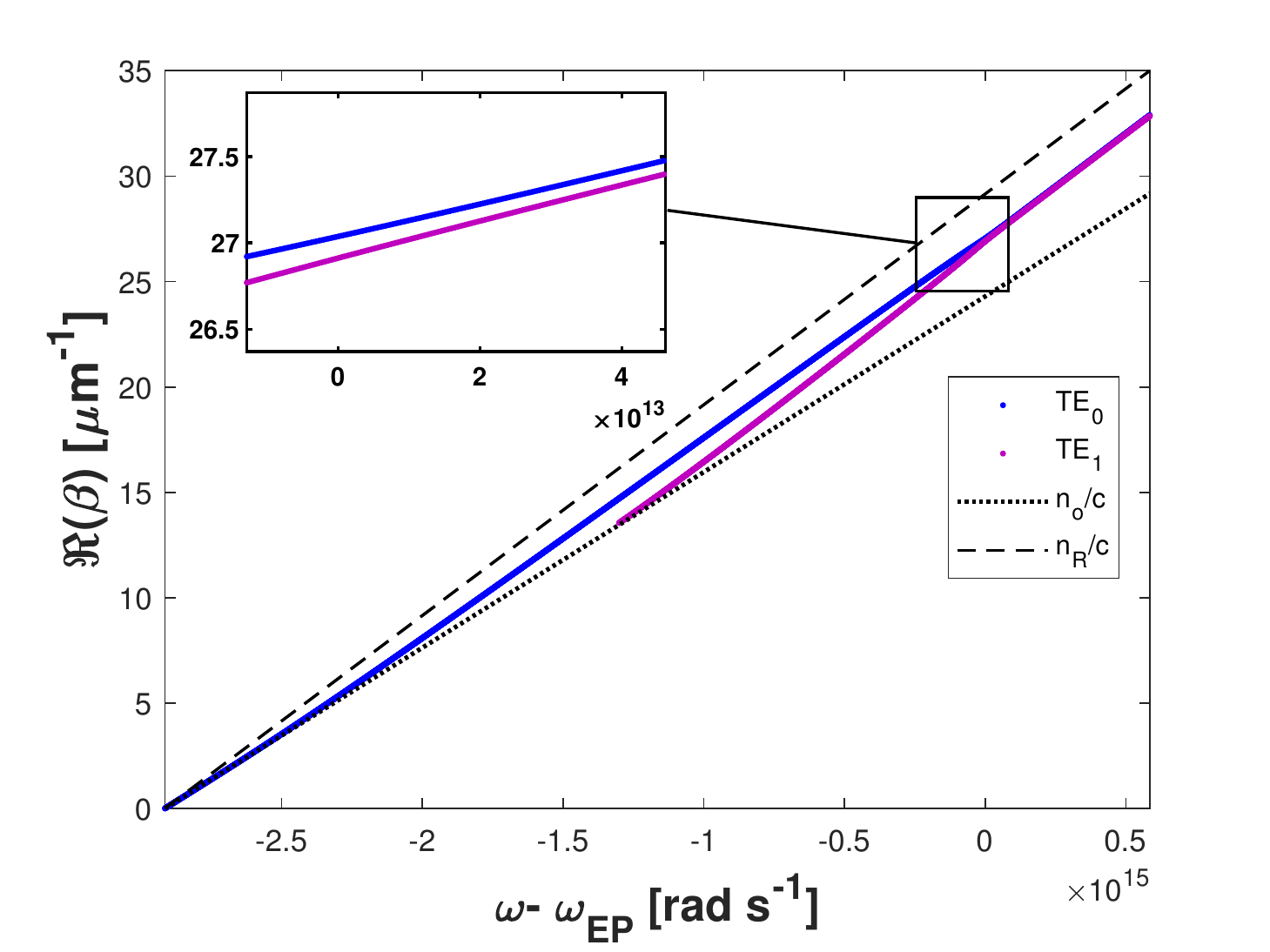}}
		\quad
	\subfloat[]
	{\includegraphics[width=0.48\textwidth]{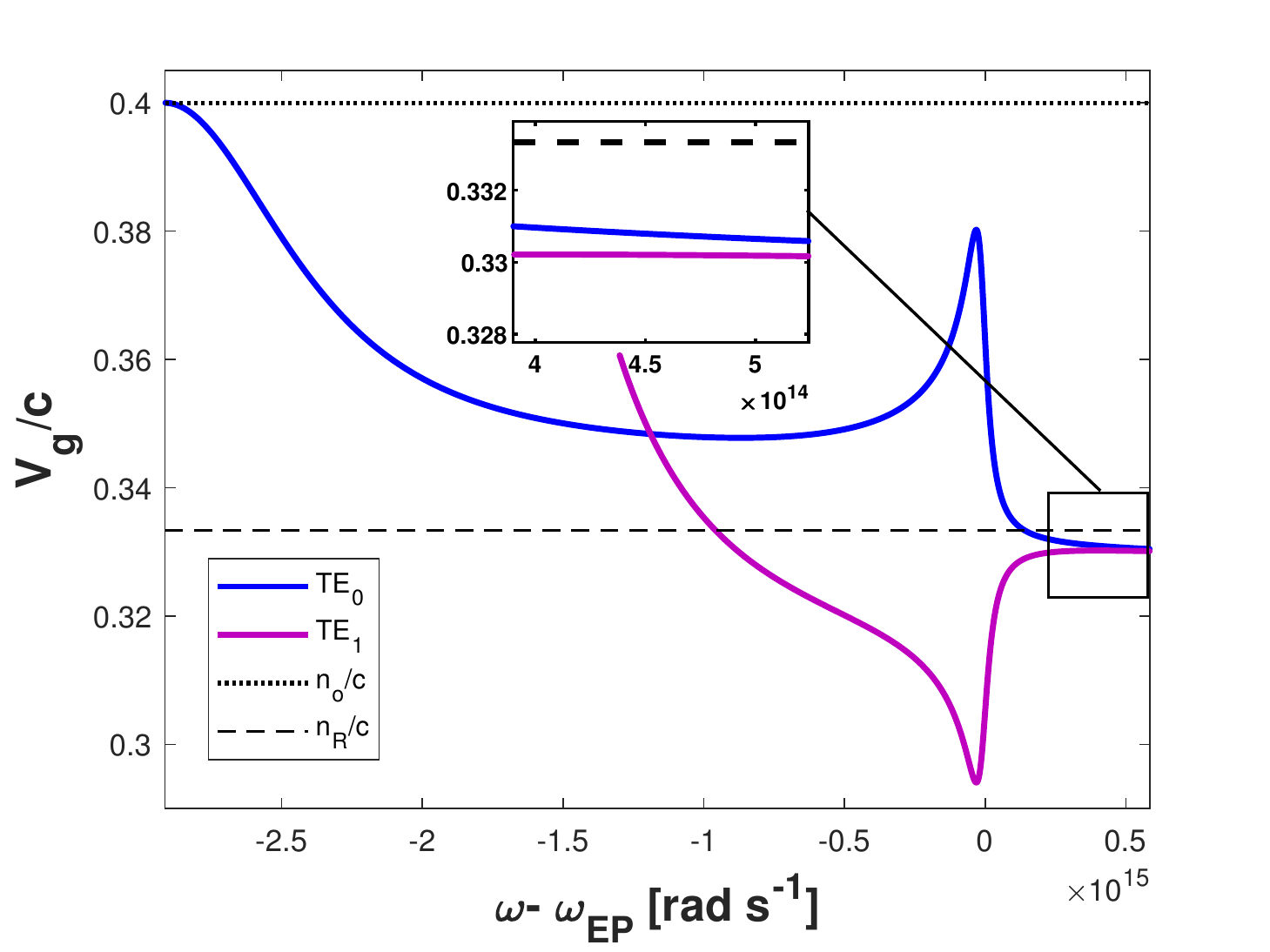}}
	\caption{(a) and (b) show the real part of the trajectories of the propagation constants as function angular frequency $\omega$. In this instance, the imbalance between the thickness of the two slabs has been increased. As a consequence, the trajectory of the group velocity of each mode does not diverge and the transition from fast to slow light does not appear anymore, as can be appreciated in (b) and (d). The numerical solutions of (a) and (b) were obtained using the parameters $n_{R}=1.6$, $n_{I}=0.02$, $n_{o}=1$, $W_{a}= 0.524 \, \mu\mathrm{m}$, $W_{b}= 0.520\, \mu\mathrm{m}$ and $d=0.524\, \mu\mathrm{m}$, meanwhile for (c) and (d) were obtained by adjusting to the set of values $n_{R}=3$, $n_{I}=0.05$, $n_{o}=2.5$, $W_{a}= 0.15 \, \mu\mathrm{m}$, $W_{b}= 0.152 \, \mu\mathrm{m}$ and $d=0.1 \, \mu\mathrm{m}$.  }
	\label{f8}
\end{figure}
In this regard, the transcendental equation \eqref{3} offers the possibility of confronting the situation when the thickness $W_{a}$ and $W_{b}$ differ slightly, bringing additional freedom in the problem of unbalancing in the gain and loss. We can therefore make use of Eq.\eqref{3}, by using the parameters in Fig.\ref{f5}, but now with the main difference that the slab thicknesses for the first waveguide configuration are $W_{a}= 0.524 \, \mu \mathrm{m}$ and $W_{b}= 0.5237 \, \mu \mathrm{m}$, whereas for the second configuration are $W_{a}= 0.15 \, \mu \mathrm{m}$ and $W_{b}= 0.151  \, \mu \mathrm{m}$. As can be seen in Figs.\ref{f7} (a) and (c), the propagation constants for the $TE_{0}$ and the $TE_{1}$ repel each other, are not equal and do not coalesce as in the perfect $\mathcal{PT}$ symmetric situation; consequently, this leads to the absence of an EP. In such a case, the transition from the broken and unbroken $\mathcal{PT}$-symmetric regions is forbidden. The supermodes'propagation constants remain complex, are anticrossing, and without a converging point. Such numerical solutions are in sharp contrast with Figs.\ref{f5} (a) and (b); specifically, the result in this section illustrates that the imperfect $\mathcal{PT}$ symmetry make the supermodes strongly asymmetric when $\beta_{TE_0} \neq \beta_{TE_1}$. In other words, there is a price to pay, which is usually attributed to the undesired detuning $\delta= \beta_{TE_1}-\beta_{TE_0}$, of their propagation constants, which stems from the unequal thicknesses of the two slabs; the propagation constants are complex and the existence of the resonant detuning tends to wash out the EP. Unlike the perfect coupling case, when both slab thickness matches, the supermodes become perfectly symmetric, with $\beta_{TE_0}=\beta_{TE_1}$, and consequently, the singularity occurs at zero-detuning. These results confirm, again, that one prerequisite for the emergence of the EP is to have a geometrically symmetric structure with a subtle balance between gain and loss. On the other hand, in the imperfect scenario, the group velocities of $TE_{0}$ and the $TE_{1}$ again display the superluminal and subluminal characters as shown in Figs.\ref{f7} (c) and (d). Nonetheless, these propagation effects become weakened when the thickness difference between the two waveguides increases until inevitably the group velocities of $TE_{0}$ and the $TE_{1}$ modes no longer experience abrupt transitions, but instead, they propagate independently, as can be appreciated in Figs.\ref{f8} (c) and (d). It is, however, worth noting from Ref.\cite{49} that one can restore the superluminal propagation and nearly reach the EP by considering a gain-and-loss unbalanced configuration in conjunction with a slight imbalance in the real part of the waveguide index. Nevertheless, it is essential to emphasize that in this case, the thicknesses of both slabs are the same. In this case, if we combine the difference in gain-loss parameters and width between the two separated slab waveguides, it could be used to expand the operating bandwidth of superluminal propagation. Such situation can be an in-depth study using the respective eigenmode analysis from our Eq.\eqref{3}.

\section{Conclusions}
In this work, we have explored analytically and numerically a $\mathcal{PT}$-symmetric two-slab system. We have based our analysis on the whole-structure model, by using the eigenmode approach to extend the results reported in Ref.\cite{48} and \cite{49}. One of the primary advantages of this formulation is that it allows for greater flexibility and control over the waveguide’s parameters, such as the background material index, the slab thickness, the separation between them, or the optical frequency. This enabled us to vary individual parameters, while keeping all others constant. As a result of the adopted formulation, a dispersion equation was derived that can be accurately solved using standard numerical routines. By numerically solving this equation, we were able to obtain the propagation constants of the $TE/TM$ eigenmodes in the waveguide structure, both for ideal perfect and imperfect $\mathcal{PT}$-symmetric configurations, being the latter accomplished from the thickness variations of both slabs. In the perfect scenario, where both guides have the same width, we have demonstrated the accuracy and usefulness of the method by reproducing previously reported examples involving only two excited modes; we have shown that all the characteristic parameters of the coupled $\mathcal{PT}$-symmetric slab waveguide can be obtained using the propagation constants of its supermodes. Moreover, the analysis revealed that the $TE$ guided modes coalesce at a certain frequency, leading to the emergence of an EP where the propagation constants associated with these modes become identical. Afterward that the propagation constants of the supermodes were found, we examined their group velocities, where the $TE_{0}$ eigenmode solution was characterized by an abrupt fast-slow light propagation within the frequency range $[\omega_{CP}, \omega_{EP} ]$. In contrast, the group velocity of the $TE_{1}$ mode is slowed down, being characteristic of slow light, within the same region without any exotic switching. Also, both group velocities tend to zero at the EP, growing monotonically after it and falling inside the range of the $V_{gmin}$ with a constant group velocity value. In the imperfect situation, for slabs with different initial thicknesses, the match between the $TE_{0}$ and $TE_{1}$ supermodes is not fulfilled and the EP is not defined. Furthermore, we find that a slight imbalance in the thickness difference between the two waveguides still allows for superluminal propagation. However, as the variation between slab thickness increases, the propagation effect is reduced until it disappears entirely. Although we have limited our solutions to the pair of $TE$ supermodes, the solutions for $TM$ supermodes can be treated similarly. Future work will address the optical force of the two eigenmodes when the system evolves between the CP and the EP regimens. In addition, the research will also focus on analyzing the leaky modes that release energy only to the substrate within the range of $\frac{\omega n_{C}}{c}< \real\left(\beta_{TE/TM}\right)<\frac{\omega n_{S}}{c}$ or for leaky modes that leak energy only to the cladding, $\frac{\omega n_{C}}{c}> \real\left(\beta_{TE/TM}\right)>\frac{\omega n_{S}}{c}$. Moreover, we planned to extend the eigenmode analysis to three coupled $\mathcal{PT}$-symmetric waveguides, where the third-order EP\textquotesingle s can emerge due to its cube-root nature; then, the rich physics of group velocity of three coupled supermodes can be analyzed and studied, which could open up opportunities to boost optical communication and computing. Lastly, we plan to publish an online simulation tool on nanoHUB, assisting students to reproduce our results and to better understand the modal diagram dispersion, propagation constants, and group velocities in a $\mathcal{PT}$-waveguide scenario.

\section{Acknowledgment}
B.M. Villegas-Martínez thanks CONACyT for financial support and the National Institute of Astrophysics, Optics and Electronics (INAOE) for supplemental support. In addition, B.M. Villegas-Martínez is grateful to Professor Demetrios Christodoulides and Midya Parto for their invaluable comments, assistance and support in this area. 

\section{Ethics declarations}
\subsection{Ethics Approval}
This material is the authors’ own original work, which has not been previously published elsewhere. All authors have been personally and actively involved in substantial work leading to the paper, and will take public responsibility for its content.

\subsection{Conflict of Interest}
The authors declare no competing interests regarding the publication of this manuscript.

\appendix 

\section{Derivation of the dispersion equation}
The transverse field components of the $TE$ and $TM$ modes of Eq.\eqref{2} satisfy the following boundary conditions at the interfaces between the $j$-th and $m$-th layer
\begin{align} \label{A1}
\phi_{j,y}\left(x\right)\big\rvert_{x=x_0}=&\phi_{m,y}\left(x\right)\big\rvert_{x=x_0}
\nonumber\\
\frac{d\phi_{j,y}\left(x\right)}{dx} \bigg\rvert_{x=x_0}=& \left(\frac{n_{j}}{n_{m}}\right)^{2p} \frac{d\phi_{m,y}\left(x\right)}{dx}\bigg\rvert_{x=x_0}, \quad \text{with} \quad p=
\begin{cases}
0 & \text{for $TE$ waves}, \\
1 & \text{for $TM$ waves}.
\end{cases}
\end{align}
Then, for the interface of region I with II, we have
\begin{align} \label{A2}
\phi_{I, y}\left(x\right)\big\rvert_{x=-a}=& \phi_{II, y}\left(x\right)\big\rvert_{x=-a}
\nonumber\\
\frac{d\phi_{I, y}\left(x\right)}{dx}\bigg\rvert_{x=-a}=& \left(\frac{n_{c}}{n_{G}}\right)^{2p} \frac{d\phi_{II, y}\left(x\right)}{dx} \bigg\rvert_{x=-a},
\end{align}
which after application in Eq.\eqref{2}, give us
\begin{align} \label{A3}
A_{1}=& A_{2} \cos\left(k_{G} W_{a}\right)-A_{3} \sin\left(k_{G}W_{a}\right), \nonumber \\
A_{1}=& \left(\frac{k_{G}}{\gamma_{C}}\right) \left(\frac{n_{C}}{n_{G}}\right)^{2p} \left[ A_{2} \sin\left(k_{G}W_{a}\right) + A_{3} \cos\left(k_{G} W_{a}\right) \right].
\end{align}
Equating these two equations yields to
\begin{equation} \label{A4}
A_{2} \left[\cos\left(k_{G} W_{a}\right)- \left(\frac{k_{G}}{\gamma_{C}}\right) \left(\frac{n_{C}}{n_{G}}\right)^{2p} \sin\left(k_{G}W_{a}\right) \right]=A_{3} \left[\left(\frac{k_{G}}{\gamma_{C}}\right) \left(\frac{n_{C}}{n_{G}}\right)^{2p}\cos\left(k_{G} W_{a}\right) + \sin\left(k_{G}W_{a}\right) \right].
\end{equation}
For the interface between II and III, we have
\begin{align} \label{A5}
\phi_{II, y}\left(x\right)\big\rvert_{x=-\frac{d}{2}} =& \phi_{III, y}\left(x\right)\big\rvert_{x=-\frac{d}{2}}
\nonumber\\
\frac{d\phi_{II, y}\left(x\right)}{dx}\bigg\rvert_{x=-\frac{d}{2}}=& \left(\frac{n_{G}}{n_{S}}\right)^{2p} \frac{d\phi_{III, y}\left(x\right)}{dx}\bigg\rvert_{x=-\frac{d}{2}},
\end{align}
which leads to
\begin{align} \label{A6}
A_{2}=&A_{4} e^{\gamma_{S} d/2} + A_{5} e^{-\gamma_{S} d/2}, \nonumber\\
A_{3}=& \left(\frac{\gamma_{S}}{k_{G}}\right) \left(\frac{n_{G}}{n_{S}}\right)^{2p} \left( A_{5} e^{-\gamma_{S} d/2} - A_{4} e^{\gamma_{S} d/2}  \right).
\end{align}
Substituting Eqs.\eqref{A6} into Eqs.\eqref{A4}, we find that
\begin{equation} \label{A7}
\frac{A_{4}}{A_{5}}=e^{-\gamma_{S}d} \left( \frac{H^{-}_{G}}{H^{+}_{G}} \right).
\end{equation}
For regions III and IV,
\begin{align} \label{A8}
\phi_{III, y}\left(x\right)\big\rvert_{x=\frac{d}{2}}=& \phi_{IV, y}\left(x\right)\big\rvert_{x=\frac{d}{2}}, 
\nonumber\\
\frac{d\phi_{III, y}\left(x\right)}{dx}\bigg\rvert_{x=\frac{d}{2}}=& \left(\frac{n_{L}}{n_{S}}\right)^{2p} \frac{d\phi_{IV, y}\left(x\right)}{dx}\bigg\rvert_{x=\frac{d}{2}},
\end{align}
and we obtain
\begin{align} \label{A9}
A_{6}=& A_{4} e^{-\gamma_{S} d/2}  + A_{5}  e^{\gamma_{S} d/2}, \nonumber \\
A_{7}=& \left(\frac{\gamma_{S}}{k_{L}}\right) \left(\frac{n_{L}}{n_{S}}\right)^{2p} \left(A_{5} e^{\gamma_{S} d/2}  - A_{4}  e^{-\gamma_{S} d/2}\right).
\end{align}
The boundary condition at the interface between IV and V are
\begin{align} \label{A10}
\phi_{IV, y}\left(x\right)\big\rvert_{x=b}=& \phi_{V, y}\left(x\right)\big\rvert_{x=b}, 
\nonumber\\
\frac{d\phi_{IV, y}\left(x\right)}{dx}\bigg\rvert_{x=b}=& \left(\frac{n_{C}}{n_{L}}\right)^{2p} \frac{d\phi_{V, y}\left(x\right)}{dx}\bigg\rvert_{x=b},
\end{align}
which give us
\begin{align} \label{A11}
A_{8}=& A_{6} \cos\left(k_{L} W_{b}\right) + A_{7} \sin\left(k_{L} W_{b}\right), \nonumber\\
A_{8}=& \left(\frac{k_{L}}{\gamma_{C}}\right) \left(\frac{n_{C}}{n_{L}}\right)^{2p} \left[A_{6} \sin\left(k_{L} W_{b}\right) - A_{7} \cos\left(k_{L} W_{b}\right) \right].
\end{align}
Equating the above equations, we find that
\begin{align} \label{A12}
A_{6}\left[\cos\left(k_{L} W_{b}\right) -\left(\frac{k_{L}}{\gamma_{C}}\right) \left(\frac{n_{C}}{n_{L}}\right)^{2p} \sin\left(k_{L} W_{b}\right)\right]=-A_{7}\left[\left(\frac{k_{L}}{\gamma_{C}}\right) \left(\frac{n_{C}}{n_{L}}\right)^{2p} \cos\left(k_{L} W_{b}\right) + \sin\left(k_{L} W_{b}\right) \right].
\end{align}
Substituting Eqs.\eqref{A9} into Eqs.\eqref{A12}, we get
\begin{equation} \label{A13}
\frac{A_{4}}{A_{5}}=e^{\gamma_{S}d}\left(\frac{H^{+}_{L}}{H^{-}_{L}}\right).
\end{equation}
Equating Eq.\eqref{A13} with Eq.\eqref{A7}, we obtain
\begin{equation} \label{A14}
\varphi \left(\beta,\omega\right)=e^{\gamma_{S}d}H^{+}_{G} H^{+}_{L}-e^{-\gamma_{S}d}H^{-}_{G} H^{-}_{L}=0.
\end{equation}

\section{Multiple EP's at different optical frequencies}
\begin{figure}[H]
    \centering
	\subfloat[]
	{\includegraphics[width=0.47\textwidth]{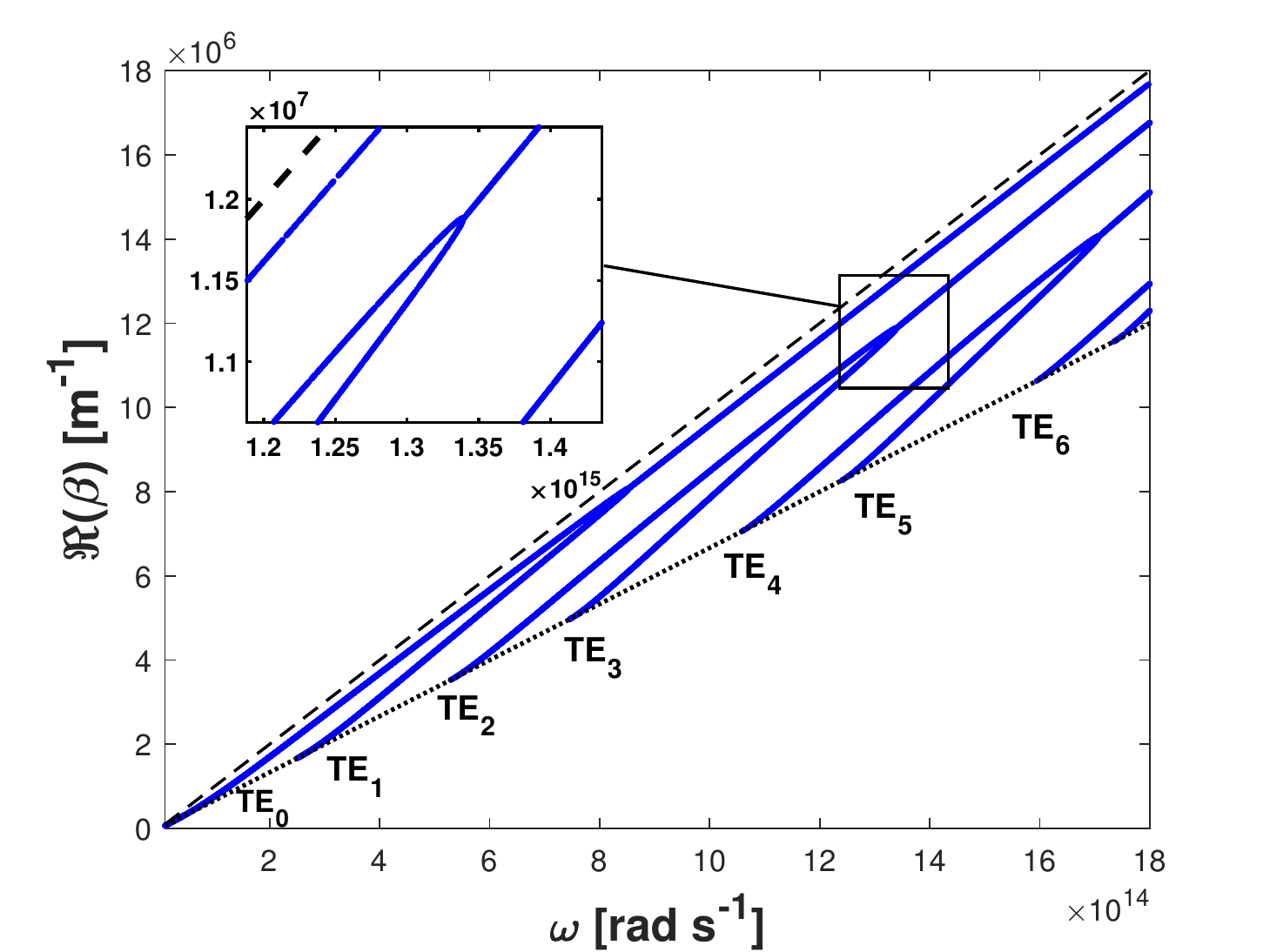}}
	\qquad
	\subfloat[]
	{\includegraphics[width=0.47\textwidth]{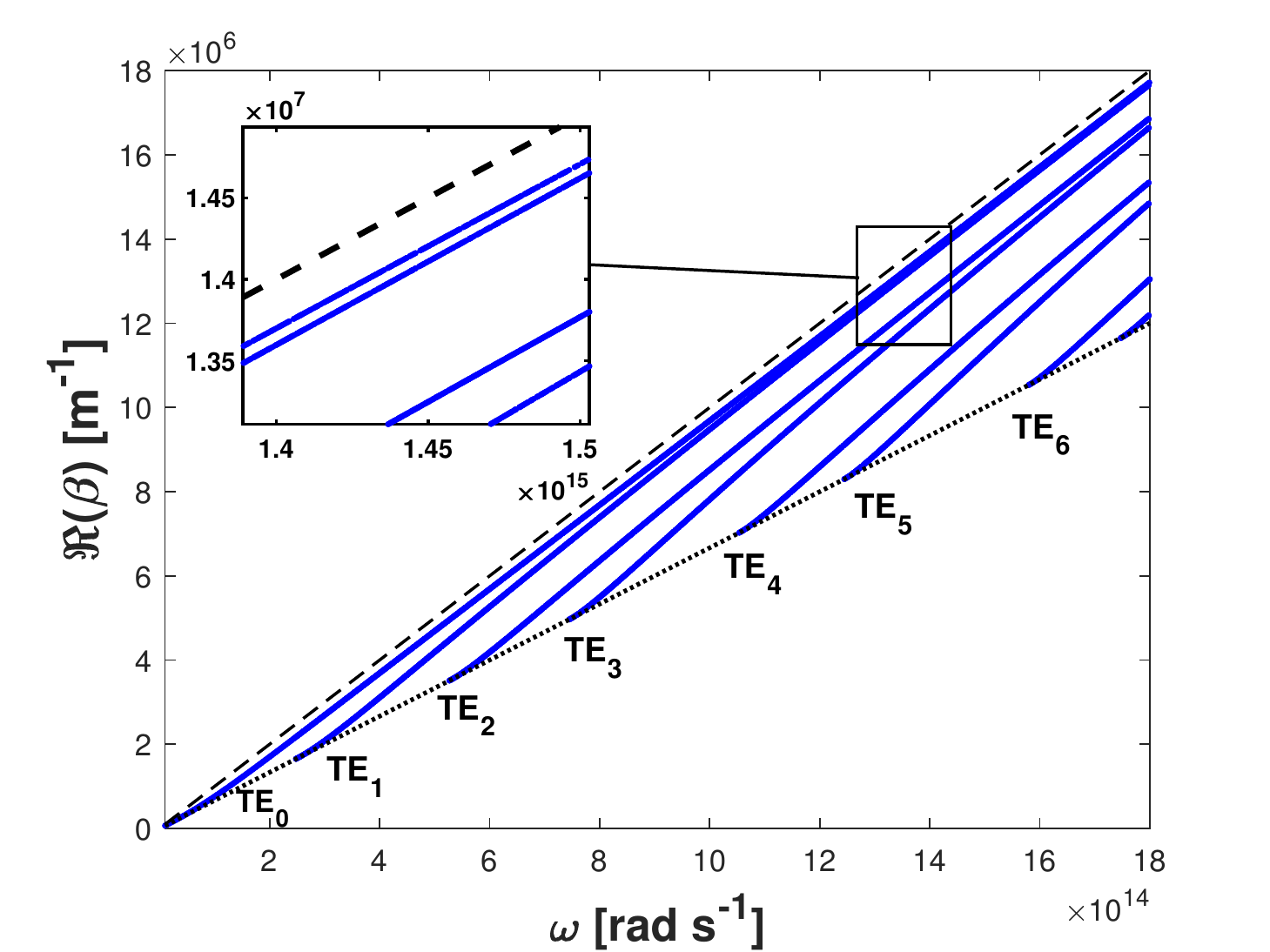}}
	\caption{The real part of the propagation constant $\beta$ versus frequency $\omega$. (a) The propagation constants of each pair of $TE$ eigenmodes remain real and opposite until they coalesce forming the EPs. At these points, the respective propagation constants become identical. A zoom on plot (a) shows the usual EP of two interacting $TE$ eigenmodes. Graph (b) outlines the dispersion curve behavior when the gain-loss parameter is not considered in the two-slab waveguide system, i.e., $\alpha=0$. As intuitively expected, the several pairs of $TE$ eigenmodes do not form the EPs, such as observed in the zoomed region. The above results were obtained using the parameters $n_{R}=3$, $n_{o}=2$, $W=0.8 \mu m$ and $d= 0.1 \mu m$.}
	\label{f9}
 \end{figure}

Since a pulse of light can be transmitted through a slab waveguide, and such slab waveguide supports several optical modes by varying their thickness, the presence of higher-order modes lead simultaneously to a major number of EP's, that could be stored whether the structure thickness is increased. In particular, the modal analysis applied here brings us the opportunity to encounter those several EP's over a broad range of optical frequencies $\omega$. To further elucidate this aspect in terms of the modal propagation constant $\beta_{TE}$, the dispersion curves for several $TE$ eigenmodes is presented in Fig.\ref{f9}.
 It is notorious that several branches of pairs $TE$ eigenmodes coalesce forming a single degenerate eigenmode at a certain frequency $\omega$, such as the one depicted in the magnified portion of Fig.\ref{f9}(a). This model can then be further analyzed to explore the hosting of multiple exceptional points. Additionally, the case where the gain-loss parameter equals zero is presented in Fig.\ref{f9}(b). These $TE$ guided modes are situated within the region demarcated by the lower, $n_{o}⁄c$, and upper cutoff frequencies, $n_{R}⁄c$, indicated by dotted and dashed lines, respectively. For practical purposes, we have omitted $TM$ solutions to easily distinguish the trajectories of the propagation constants of $TE$ eigenmodes.

\section{Full-field solution}
Once the dispersion equation has been found, we may set freely any amplitude coefficient to unity; in this case, and without loss of generality, we fix $A_{8} = 1$. Then, we can now easily find the amplitude coefficients $A_{j}$ with $j = 1$ through 7; in this case, we use the Eq.\eqref{A4} into the first expression of \eqref{A3} to obtain
\begin{align} \label{B1}
A_{1}=&A_{2} \cos\left(k_{G} W_{a} \right) \left[1-\epsilon_{G}\tan\left(k_{G} W_{a}\right)\right], \nonumber\\
A_{3}=&\epsilon_{G} A_{2},
\end{align}
with
\begin{equation} \label{B2}
\epsilon_{G}=\frac{\cos\left(k_{G} W_{a}\right)-\left(\frac{k_{G}}{\gamma_{C}}\right)\left(\frac{n_{C}}{n_{G}}\right)^{2q}\sin\left(k_{G} W_{a}\right)}{\left(\frac{k_{G}}{\gamma_{C}}\right)\left(\frac{n_{C}}{n_{G}}\right)^{2q}\cos\left(k_{G} W_{a}\right) + \sin\left(k_{G} W_{a}\right) }.
\end{equation}
For $A_{4}$ and $A_{5}$, we use Eq.\eqref{A7} into the first expression of \eqref{A6}, to get 
\begin{align}  \label{B3}
A_{4}=&A_{2} e^{-\gamma_{S}d/2} \left( \frac{H^{-}_{G}}{H^{+}_{G} + H^{-}_{G}} \right), \nonumber\\
A_{5}=&A_{2} e^{\gamma_{S}d/2} \left( \frac{H^{+}_{G}}{H^{+}_{G} + H^{-}_{G}} \right).
\end{align}
Substituting the expressions for $A_{4}$ and $A_{5}$ above into Eq.\eqref{A9},
\begin{align}  \label{B4}
A_{6}=& \frac{A_{2}}{H^{+}_{G} + H^{-}_{G}}  \left( H^{+}_{G} e^{\gamma_{S}d} + H^{-}_{G} e^{-\gamma_{S}d} \right) , \nonumber\\
A_{7}=& \left(\frac{\gamma_{S}}{k_{L}}\right) \left(\frac{n_{L}}{n_{S}}\right)^{2q}  \frac{A_{2}}{H^{+}_{G} + H^{-}_{G}}  \left( H^{+}_{G} e^{\gamma_{S}d} - H^{-}_{G} e^{-\gamma_{S}d} \right) .
\end{align}
Finally, we find the amplitude coefficient $A_{2}$ by using the result of $A_{6}$ in conjunction with Eq.\eqref{A12} into the first expression of Eq.\eqref{A11}, 
\begin{equation} \label{B5}
A_{2}=\frac{H^{+}_{G} + H^{-}_{G}}{\left(H^{+}_{G} e^{\gamma_{S}d} + H^{-}_{G} e^{-\gamma_{S}d}\right) \cos\left(k_{L} W_{b} \right) \left[1-\epsilon_{L}\tan\left(k_{L} W_{b}\right)\right]},
\end{equation}
where
\begin{equation} \label{B6}
\epsilon_{L}=\frac{\cos\left(k_{L} W_{b}\right)-\left(\frac{k_{L}}{\gamma_{C}}\right)\left(\frac{n_{C}}{n_{L}}\right)^{2q}\sin\left(k_{L} W_{b}\right)}{\left(\frac{k_{L}}{\gamma_{C}}\right)\left(\frac{n_{C}}{n_{L}}\right)^{2q}\cos\left(k_{L} W_{b}\right) + \sin\left(k_{L} W_{b}\right) }.
\end{equation}

\end{document}